\documentclass[prx,twocolumn,superscriptaddress,longbibliography]{revtex4-2}

\usepackage[dvipdfmx]{graphicx}
\usepackage{graphicx}
\usepackage{amsmath,amssymb,amsthm,mathrsfs,amsfonts,dsfont}
\usepackage{subfigure, epsfig}
\usepackage{braket}
\usepackage{bm}
\usepackage{enumerate}
\usepackage{comment}
\usepackage[colorlinks,linkcolor=blue,citecolor=blue]{hyperref}
\newcommand{\tr}{\mathrm{Tr}}

\newcommand{\mathbbm}[1]{\text{\usefont{U}{bbm}{m}{n}#1}} 
\usepackage[normalem]{ulem} 
\usepackage{bibunits}
\usepackage{appendix}

\defaultbibliographystyle{apsrev4-2}

\begin{document}
\title{Experimentally probing entropy reduction via iterative quantum information transfer}

\author{Toshihiro Yada}
\thanks{These authors contributed equally to this work.}
\affiliation{Department of Applied Physics, The University of Tokyo, 7-3-1 Hongo, Bunkyo-ku, Tokyo 113-8656, Japan}

\author{Pieter-Jan Stas}
\thanks{These authors contributed equally to this work.}
\affiliation{Department of Physics, Harvard University, Cambridge, Massachusetts 02138, USA}

\author{Aziza Suleymanzade}
\affiliation{Department of Physics, Harvard University, Cambridge, Massachusetts 02138, USA}

\author{Erik N. Knall}
\affiliation{John A. Paulson School of Engineering and Applied Sciences, Harvard University, Cambridge, Massachusetts 02138, USA}

\author{Nobuyuki Yoshioka}
\affiliation{International Center for Elementary Particle Physics, The University of Tokyo, 7-3-1 Hongo, Bunkyo-ku, Tokyo 113-0033, Japan}
\affiliation{JST, PRESTO, 4-1-8 Honcho, Kawaguchi, Saitama, 332-0012, Japan}

\author{Takahiro Sagawa}
\affiliation{Department of Applied Physics, The University of Tokyo, 7-3-1 Hongo, Bunkyo-ku, Tokyo 113-8656, Japan}
\affiliation{Quantum-Phase Electronics Center (QPEC), The University of Tokyo,7-3-1 Hongo, Bunkyo-ku, Tokyo 113-8656, Japan}

\author{Mikhail D. Lukin}
\affiliation{Department of Physics, Harvard University, Cambridge, Massachusetts 02138, USA}


\begin{abstract}
Thermodynamic principles governing energy and information are important tools for a deeper understanding and better control of quantum systems. In this work, we experimentally investigate the interplay of the thermodynamic costs and information flow in a quantum system undergoing iterative quantum measurement and feedback. Our study employs a state stabilization protocol involving repeated measurement and feedback on an electronic spin qubit associated with a Silicon-Vacancy center in diamond, which is strongly coupled to a diamond nanocavity. This setup allows us to verify the fundamental laws of nonequilibrium quantum thermodynamics, including the second law and the fluctuation theorem, both of which incorporate measures of quantum information flow induced by iterative measurement and feedback. We further assess the reducible entropy based on the feedback's causal structure and quantitatively demonstrate the thermodynamic advantages of non-Markovian feedback over Markovian feedback. For that purpose, we extend the theoretical framework of quantum thermodynamics to include the causal structure of the applied feedback protocol. Our work lays the foundation for investigating the entropic and energetic costs of real-time quantum control in various quantum systems.
\end{abstract}
\maketitle

\section{Introduction}
Recent advances in nonequilibrium thermodynamics \cite{PRXinfothermoExp,esposito2009nonequilibrium,campisi2011colloquium,seifert2012stochastic,funo2018quantum,landi2021irreversible} have revealed the fundamental relationship between information and thermodynamics \cite{leff2002maxwell,parrondo2015thermodynamics}.
The reduction of entropy in a system through feedback control by ``Maxwell's demon" is fundamentally bounded by the information accessible to the demon, which is characterized by mutual information and information flow. 
This is reflected in generalizations of the second law of thermodynamics (SL) and the fluctuation theorem (FT) \cite{sagawa2008second,sagawa2010generalized,horowitz2010nonequilibrium,sagawa2012nonequilibrium,ito2013information,hartich2014stochastic,HorowitzEspositoPRX2014,funo2013integral,ptaszynski2019thermodynamics,yada2022quantum,strasberg2017quantum,strasberg2019operational}, in which information and thermodynamic quantities are treated on an equal footing even in far from equilibrium dynamics. In the classical regime, the generalized thermodynamic principles have been experimentally verified, clarifying the role of the continuous flow of classical information \cite{ribezzi2019large,debiossac2020thermodynamics,debiossac2022non}, beyond single measurement-feedback protocols \cite{toyabe2010experimental,berut2012experimental,roldan2014universal,koski2014experimental,vidrighin2016photonic,chida2017power,paneru2018lossless,paneru2020efficiency,yan2024experimental}. In the quantum regime \cite{camati2016experimental,cottet2017observing,masuyama2018information,naghiloo2018information,solfanelli2021,hernandez2022autonomous}, however, the thermodynamic role of information flow induced by iterative quantum feedback has not been probed experimentally.

Exploring the impact of information flow on quantum thermodynamics is critical to characterize the dynamical behavior of a system under multiple rounds of quantum measurement and feedback, which is practically relevant to various tasks in quantum experiments. Iterative or continuous feedback control can be used to create specific quantum resource states \cite{sayrin2011real,magrini2021real} and to execute quantum error correction protocols \cite{ofek2016extending,ryan2021realization,krinner2022realizing}. 
To perform these tasks effectively, it is crucial to find the optimal feedback protocol that fully exploits the available information \cite{wiseman2009quantum,sivak2022model}. However, the thermodynamic description of the causal structure of quantum control remains to be investigated, unlike the well-explored classical case \cite{horowitz2010nonequilibrium,sagawa2012nonequilibrium,ito2013information,hartich2014stochastic,HorowitzEspositoPRX2014,horowitz2014second,barato2014efficiency,hartich2016sensory,ito2016backward}.

\begin{figure*}[]
\begin{center}
\includegraphics[width=0.8\textwidth]{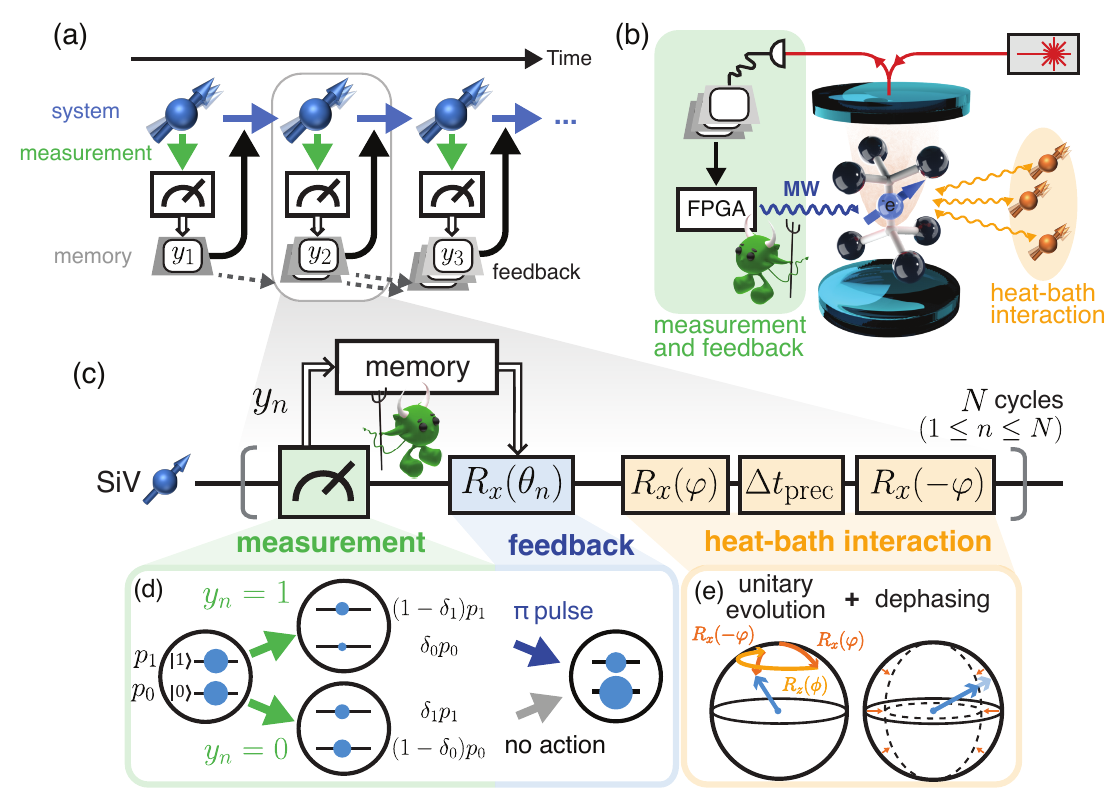}
\caption{The Silicon-vacancy center under iterative measurement and feedback. (a) Schematic of a quantum system under repeated measurement and feedback. The operation in the $n$-th feedback process is determined by all past measurement outcomes $(y_1,y_2,\dots,y_n)$. (b) Experimental setup: The SiV coupled to a nanophotonic optical cavity in diamond is placed in a cryostat at a $200 \mathrm{mK}$ temperature. A readout laser at $\sim 737 \mathrm{nm}$ is sent to the SiV-cavity system through an optical fiber, and the reflected light intensity is monitored with a superconducting nanowire single-photon detector to determine the SiV electron spin state. The measurement results are stored and sent to an FPGA that plays the role of Maxwell's demon and can apply quantum gates on the SiV electronic spin qubit through the MW application. The decoherence mechanism is established through SiV's interaction with a heat bath consisting of spins in the diamond lattice. (c) A detailed experimental sequence of a step shown in (a): The SiV electronic spin qubit is measured, and the result $y_n$ is sent to the Maxwell's demon's memory, which then applies the feedback unitary rotation $\hat{R}_x(\theta_n)$ based on the past result(s). The spin then evolves for time $\Delta t_{\mathrm{prec}}$ in a tilted basis set by the pre- and post-rotation gates $\hat{R}_x(\varphi)$ and $\hat{R}_x(-\varphi)$. (d) Details of the measurement and feedback with a specific Markovian feedback. The population change by the measurement and feedback is illustrated separately for $y_n=1$ and $y_n=0$, where coherence between $\ket{0}$ and $\ket{1}$ is ignored in this figure for simplicity. (e) Bloch sphere representation of the unitary evolution and dephasing.}
\label{fig:setup_and_sequence}
\end{center}
\end{figure*}

In this work, we experimentally explore the role of information flow in quantum thermodynamics using successive quantum feedback on a solid-state spin qubit.
Specifically, our quantum system consists of a Silicon-Vacancy (SiV) defect center coupled to a diamond nanocavity \cite{nguyen2019integrated,stas2022}.
Using our platform, we first experimentally validate the fundamental thermodynamic laws, the SL and FT, which incorporate a measure of quantum information flow, as theoretically predicted in Ref.~\cite{yada2022quantum}. These results confirm that the measure of information flow bounds the fundamental capability of quantum feedback control, i.e., the amount of entropy reduction.
Moreover, we experimentally explore the role of the causal structure of the feedback protocol, which is responsible for how efficiently the obtained information is used.  By turning on and off the dependence of the feedback on the past measurement outcomes, we generate non-Markovian and Markovian control sequences as different feedback causal structures.
To properly describe the effect of non-Markovian feedback protocols, we expand the theoretical framework \cite{yada2022quantum} and derive the generalized SL and FT by quantifying the unavailable information due to the constraints on feedback design.

Based on the theoretical framework developed in this study, we experimentally show the increased accessible information and reducible entropy of non-Markovian feedback over Markovian feedback, thus not only highlighting but also quantifying the benefit of non-Markovianity in thermodynamic terms. Our framework offers a new toolset for understanding the role of information flow, optimizing feedback to control quantum systems, learning the energy cost of quantum information processing, and designing efficient quantum heat engines under feedback control in the real-time domain.

\section{Key idea: thermodynamics under feedback}

Originally developed for macroscopic systems, thermodynamics has since been extended to microscopic systems where thermal and quantum fluctuations play a significant role \cite{esposito2009nonequilibrium,campisi2011colloquium}. Due to these fluctuations, physical systems follow different trajectories for individual trials, and stochastic thermodynamic quantities can be defined for each trajectory. For example, in an ensemble of orthogonal basis $\{\ket{x}\}_x$ with probability $p(x)$, the stochastic entropy for $\ket{x}$ can be defined as $s \equiv -\ln p(x)$, whose average denoted by $\langle \cdot \rangle$ becomes the von Neumann entropy of $\rho \equiv \sum_{x} p(x) \ket{x}\bra{x}$ (i.e., $\langle s \rangle = S(\rho) \equiv -\tr[\rho \ln \rho]$).

The SL and FT are the fundamental thermodynamic principles of such stochastic systems. The SL is denoted as $\langle \sigma \rangle \geq 0$ with the entropy production defined as $\sigma \equiv s_f - s_i - \beta Q$ \cite{deffner2011nonequilibrium,horowitz2013entropy,hekking2013quantum}, where $s_f$ and $s_i$ represent the stochastic entropy of the final and initial states, $\beta$ is the inverse temperature of the heat bath, and $Q$ denotes the input heat from the bath. The entropy production quantifies the thermodynamic cost required for quantum dynamics \cite{landi2021irreversible}.
While the SL only concerns the average value of the entropy production $\langle \sigma \rangle$, the FT, defined as $\langle e^{-\sigma} \rangle =1$, also includes its higher-order moments, thus describing the system's thermodynamic behavior at the level of fluctuations. The FT plays a central role in nonequilibrium thermodynamics since it leads to the derivation of many thermodynamic laws, including the SL and the fluctuation-dissipation theorems \cite{andrieux2004fluctuation,saito2008symmetry}.

In the presence of feedback by Maxwell's demon, the entropy production can be reduced by the amount of the demon's information gain $i$, resulting in the generalization of the SL as $\langle \sigma \rangle \geq -\langle i\rangle$ and the FT as $\langle e^{-\sigma-i}\rangle = 1$.
For classical systems, the information $i$ is typically characterized by the transfer entropy \cite{sagawa2012nonequilibrium,ito2013information,hartich2014stochastic}. It is a well-established measure of classical information flow \cite{schreiber2000measuring}, widely used in time series analysis across various fields, including finance \cite{granger1969investigating}, neuroscience \cite{vicente2011transfer}, and biochemistry \cite{ito2015maxwell}.

For a quantum system under iterative measurement and feedback, shown in Fig.~\ref{fig:setup_and_sequence}(a), the relevant measure of information flow is the quantum-classical-transfer (QC-transfer) entropy \cite{yada2022quantum}, denoted as $i_{\rm QCT}$, and the SL and the FT are generalized as follows (see Appendix~\ref{apps:thermo_QCTE} for details):
\begin{gather}
 \langle \sigma \rangle \geq  -\langle i_{\rm QCT}\rangle, \label{eq:GSL} \\
    \langle e^{-\sigma-i_{\rm QCT}} \rangle = 1. \label{eq:GFT}
\end{gather}
The QC-transfer entropy $i_{\rm QCT}$ is defined as the accumulation of newly obtained information in each measurement, with its increment in the $n$-th measurement process given by
\begin{gather}
    \Delta i_{\rm QCT} = -\ln p(b_n|Y_{n-1}) + \ln p(a_n|Y_{n}), \label{eq:t_QCT} \\
    \Delta \langle i_{\rm QCT} \rangle = S(\rho_n|Y_{n-1}) - S(\tau_n|Y_{n}).  \label{eq:e_QCT}
\end{gather}
Here, the $n$-th measurement outcome is expressed as $y_n$, and all the outcomes until $y_n$ are expressed as $Y_n\equiv (y_1,y_2,\dots,y_n)$. 
The conditional probabilities $p(b_n|Y_{n-1})$ and $p(a_n|Y_n)$ correspond to the populations of the conditional density operators $\rho_n^{Y_{n-1}}$ and $\tau_n^{Y_n}$ in their respective diagonal bases. These operators represent the system's quantum states immediately before and after the $n$-th measurement, respectively. The indices $b_n$ and $a_n$ label the eigenstates of those diagonalizations, realized in each quantum trajectory.
$\Delta \langle i_{\rm QCT} \rangle$ quantifies how much the conditional entropy $S(\rho|y) \equiv \sum_y  p_y S(\rho^y)$ is reduced through the $n$-th measurement process.

Our experiments explore the dynamics of a single quantum spin system under repeated measurement and feedback (Fig.~\ref{fig:setup_and_sequence}(b)). We extract thermodynamic quantities such as entropy and information gained from measurements to verify and study the thermodynamic laws. Applying weak measurements and implementing both coherent and incoherent quantum evolution in between successive measurement steps allows us to probe regimes of nontrivial evolution of thermodynamic quantities. The outcomes allow us to explore the differences between quantum and classical dynamics and properties of the  information flow that depend on the causal structure of implemented feedback protocols.

\section{Experimental system}

We treat the electron spin degree of freedom of the SiV center in a diamond nanocavity in our experimental setup as a two-level system or a qubit with the ground $\ket{0}$ (corresponding to $\ket{\downarrow}$) and excited $\ket{1}$ (corresponding to $\ket{\uparrow}$) states split by an externally applied magnetic field. We iteratively measure the state of the spin qubit in the $\hat{\sigma}_z$-basis. The measurement outcomes are then fed back into the system through the application of microwave (MW) pulses near-resonant ($\Omega_{\omega} \gg \Delta_\omega$, with $\Omega_{\omega}$ and $\Delta_\omega$ the pulse Rabi frequency and detuning, respectively) with the spin qubit MW transition to stabilize either the ground or excited state.
The qubit is then left to evolve for time $\Delta t_{\mathrm{prec}}$ under a constant Hamiltonian while also experiencing dephasing noise through the interaction with the $^{13}$C nuclear spin bath in diamond. The $^{13}$C nuclear spins are modeled as a heat bath whose inverse temperature is $\beta = 0$.
For all experiments, we initialize the qubit in a maximally mixed state so that the success of the target state stabilization implies the entropy reduction $\langle \sigma \rangle =\langle s_f\rangle-\langle s_i \rangle <0$.

The experimental protocol shown in Fig.~\ref{fig:setup_and_sequence}(c) consists of successive rounds of measurement, feedback, and heat-bath interactions.
We measure the qubit by shining a readout laser near the SiV electron resonance and monitoring the reflected photon counts to determine the qubit state. This measurement process can be described by the positive operator-valued measure (POVM)with operators $E_{y_n=0} \equiv (1-\delta_0)\ket{0}\bra{0} + \delta_1\ket{1}\bra{1}$ and $E_{y_n=1} \equiv \delta_0\ket{0}\bra{0} + (1-\delta_1)\ket{1}\bra{1}$, where the parameter $\delta_{0}$ (resp. $\delta_1$) represents the error probability that $y_n=1$ (resp. $y_n=0$) is read out, even though the system is in the $\ket{0}$ (resp. $\ket{1}$) state. The values of $\delta_{0}$ and $\delta_1$ are adjusted by setting the readout time and tuning the laser frequency with respect to the SiV optical transition, which enables us to explore different information transfer regimes. 
Details of the optical readout, including the definition and tuning of $\delta_0$ and $\delta_1$, are provided in Appendix~\ref{apps:opt_readout}.

The measurement outcomes are immediately processed by a Field Programmable Gate Array (FPGA), and fed back in $50\mathrm{ns}$, which is much shorter than the qubit coherence time. In each feedback process, a $\theta_n$ rotation around the $x$-axis $\hat{R}_x(\theta_n)\equiv e^{-i\theta_n\hat{\sigma}_x}$ is applied with a MW pulse, where $\theta_n$ depends on past measurement outcome(s) through a pre-programmed feedback function.

The system then experiences noise against the stabilization of the $\hat{\sigma}_z$ eigenstates (Fig. ~\ref{fig:setup_and_sequence}(e)) through dephasing ($T^*_2 \sim 4 \mathrm{\mu s}$) as well as coherent evolution under the Hamiltonian $\hat{H} = \Delta_{\omega}\hat{\sigma}_z$ for duration $\Delta t_{\mathrm{prec}}$, where $\Delta_{\omega}$ is the detuning between the MW drive frequency and the qubit MW transition frequency. As $T_1> 1 \mathrm{s}$ for the SiV, the effect of $T_1$-dependent depolarization is not significant and can be ignored. Since both the dephasing and coherent evolution processes leave the $\hat{\sigma}_z$ eigenstates unaffected, we apply rotations $\hat{R}_x(\varphi)$ and $\hat{R}_x(-\varphi)$ before and after the evolution time $\Delta t_{\mathrm{prec}}$ to effectively implement off-axis (from the measurement basis) evolution of the system. The coherent evolution can then be described as the application of the unitary gate $\hat{V}\equiv \hat{R}_x(\varphi)\hat{R}_z(\phi)\hat{R}_x(-\varphi)$, where $\phi = \Delta_{\omega}\Delta t_{\mathrm{prec}}$.
The unitary gate $\hat{V}$ introduces the quantum nature into the system's dynamics by generating quantum coherence between $\ket{0}$ and $\ket{1}$. Without $\hat{V}$, the dynamics of the system would be completely classical as a mixture of $\ket{0}$ and $\ket{1}$ states, and thus not suitable for our exploration of thermodynamics in the quantum regime. Additional details about experimental parameters and analysis can be found in Sec. S1 of the Supplemental Material.

\section{Experimental Results}

\subsection{Verification of the generalized thermodynamic laws under Markovian feedback}\label{ss:Markov_exp}

\begin{figure}[]
\begin{center}
\includegraphics[width=0.45\textwidth]{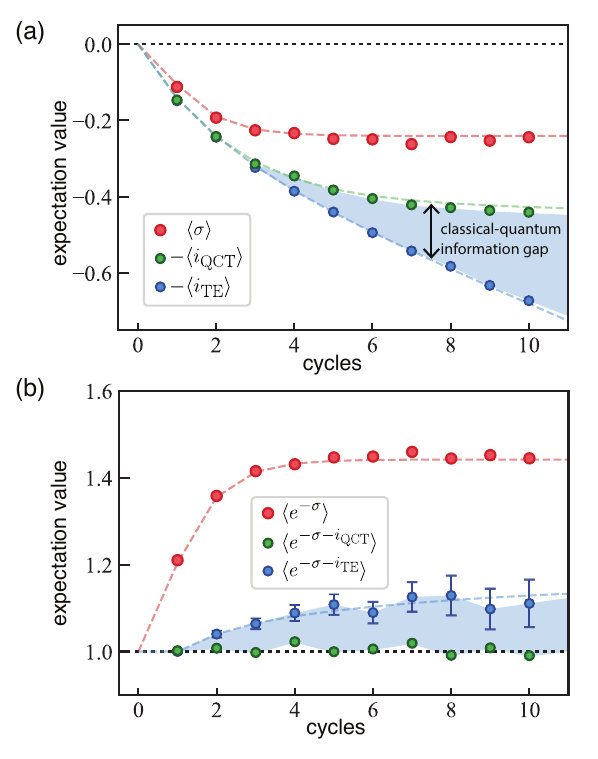}
\caption{Evolution of thermodynamic quantities under Markovian feedback. (a) Experimentally obtained entropy production $\langle \sigma \rangle$, negative transfer entropy $-\langle i_{\rm TE}\rangle$, and negative QC-transfer entropy $-\langle i_{\rm QCT}\rangle$ against the number of feedback cycles. The blue shaded region illustrates the information loss due to measurement back action, indicating the gap between $-\langle i_{\rm TE}\rangle$ and $-\langle i_{\rm QCT}\rangle$. (b), Statistical averages $\langle e^{-\sigma}\rangle$, $\langle e^{-\sigma -i_{\rm TE}}\rangle$, and $\langle e^{-\sigma -i_{\rm QCT}}\rangle$ from experimental data are plotted against the number of feedback cycles. The blue shaded region highlights the gap between the ensemble averages with classical information measure $\langle e^{-\sigma -i_{\rm TE}}\rangle$ and quantum information measure $\langle e^{-\sigma -i_{\rm QCT}}\rangle$. In both (a) and (b), error bars, which represent the standard error of the mean, are smaller than the marker sizes except for $\langle e^{-\sigma -i_{\rm TE}}\rangle$. 
Dotted lines represent theoretical predictions for the same colored plots, obtained from numerical simulations using the experimental parameters.}
\label{fig:MarkovExp}
\end{center}
\end{figure}

We begin by experimentally verifying the generalized SL (\ref{eq:GSL}) and FT (\ref{eq:GFT}) in our system by applying a Markovian feedback protocol. The applied sequence, shown in Fig.~\ref{fig:setup_and_sequence}(d), is Markovian because it does not utilize the memory of past measurements. The resulting feedback pulse angle $\theta_n$ depends on the measurement result $y_n$ only: $\theta_n=\pi$ if $y_n=1$ and $\theta_n=0$ if $y_n=0$. This effectively stabilizes the ground state $\ket{0}$. The applied sequence contains 10 successive cycles of measurement and feedback. 
To obtain the thermodynamic quantities shown in Fig.~\ref{fig:MarkovExp}, we repeated the experimental trial 7000 times and computed the ensemble averages over the sampled trajectories.

Fig.~\ref{fig:MarkovExp}(a) shows a reduction of the entropy of the system (i.e., $\langle \sigma \rangle < 0$) due to the ground-state stabilization effect of the feedback sequence. At the same time, the entropy is lower-bounded by the negative QC-transfer entropy $-\langle i_{\rm QCT} \rangle$ in accordance with the generalized SL (\ref{eq:GSL}). The observed gap between $\langle \sigma \rangle$ and $-\langle i_{\rm QCT} \rangle$ reflects the fact that not all of the extracted information is used for entropy reduction, primarily due to the Markovian nature of the feedback protocol, which ignores the past outcomes.
We also observe that the classical transfer entropy $i_{\rm TE}$, while a well-established measure of classical information flow \cite{schreiber2000measuring}, is not a suitable measure in the quantum regime. The increment of $i_{\rm TE}$ in the $n$-th measurement process, given by $\Delta i_{\rm TE} = -\ln p(b_n|Y_{n-1}) + \ln p(b_n|Y_n)$, provides a looser bound on the entropy reduction than its quantum counterpart $i_{\rm QCT}$. This is because $i_{\rm TE}$ does not capture the state change induced by measurement backaction, whereas $i_{\rm QCT}$ does. Further details on this distinction are provided in Appendix~\ref{Appendix:quantum_classical}.

Fig.~\ref{fig:MarkovExp}(b) shows that the generalized FT with QC-transfer entropy (\ref{eq:GFT}) holds, with $\langle e^{-\sigma -i_{\rm QCT}}\rangle \sim 1$ throughout all measurement and feedback cycles. We observe that the uncorrected statistical average $\langle e^{-\sigma}\rangle$ deviates from 1 due to the presence of feedback control, mirroring the negative entropy production shown in Fig.~\ref{fig:MarkovExp}(a). We can also observe that the classical transfer entropy $i_{\rm TE}$ does not accurately describe the information flow in this experiment as the generalized FT with classical transfer entropy $i_{\rm TE}$ clearly does not hold, and $\langle e^{-\sigma-i_{\rm TE}}\rangle$ significantly diverges from 1. We note that since $i_{\rm QCT}$ reduces to $i_{\rm TE}$ under fully classical dynamics, these divergences can also be interpreted as indicating the quantum nature of the system's dynamics (see Appendix~\ref{Appendix:quantum_classical} for more information on the relationship between quantum and classical information measures).
In both Figs.~\ref{fig:MarkovExp}(a) and (b), experimental measurements of the thermodynamic quantities are in good agreement with the theoretical predictions. Our results confirm that the QC-transfer entropy is a relevant information measure under iterative quantum feedback, which characterizes thermodynamic quantities both at the ensemble and individual trajectory levels.

\subsection{Non-Markovian feedback and the role of causal structure of feedback in quantum thermodynamics}

\begin{figure*}[]
\begin{center}
\includegraphics[width=1\textwidth]{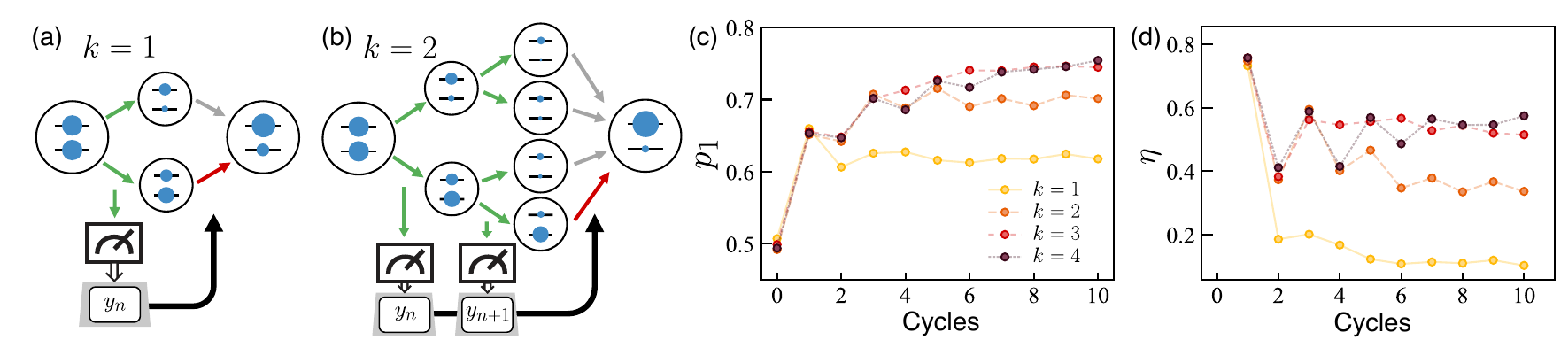}
\caption{$k$-th-order Markovian feedback protocols and their performance in the experiments. (a),(b) Schematics for Markovian and second-order Markovian feedback protocols stabilizing the excited state $\ket{1}$. The free evolution between the measurement of $y_n$ and $y_{n+1}$ is not shown in (b) for simplicity. (c) Experimentally measured excited-state populations $p_1$ under $k$-th-order Markovian feedback ($k=1,2,3,4$) as a function of the number of feedback cycles. (d) Experimentally measured conversion efficiencies for the same $k$-th-order Markovian feedback conditions ($k=1,2,3,4$) plotted against the number of feedback cycles. As shown, the non-Markovianity enhances the information-thermodynamic efficiency of the feedback protocol.}
\label{fig:nonMarkovExp}
\end{center}
\end{figure*}

While the above section treated a memory-less (that is, Markovian) feedback protocol, many feedback protocols contain a more general causal structure incorporating multiple past measurements stored in memory to make an optimum feedback decision. Systems under such non-Markovian feedback are best described by non-Markovian dynamics.
In the following, we first extend the theoretical framework to quantify how much information is actually used for the entropy reduction, depending on the causal structure (or Markovianity) of feedback.
Based on that, we experimentally demonstrate and measure the thermodynamic advantage of non-Markovian feedback over Markovian feedback.

In order to formulate the causal structure of feedback, we introduce the notion of \textit{$k$-th-order Markovian} feedback, in which only the last $k$ measurement outcomes are utilized in designing the feedback protocol. 
The entropy reduction achieved with $k$-th-order Markovian feedback can be characterized in a manner that reflects the restriction imposed on the feedback design.
To this end, we introduce the information measure $i_{\rm FB}^k$, which quantifies the net amount of information that can be fed back to the system if a $k$-th-order Markovian protocol is employed (see Appendix~\ref{apps:Qthermo_BQC} for the precise definition).
Since measurement outcomes other than the most recent $k$ are not used in such a protocol, the inequality $\langle i_{\rm FB}^{k} \rangle \leq \langle i_{\rm QCT} \rangle$ holds, with $\langle i_{\rm QCT} \rangle$ being the total information gain available from all measurement outcomes.
Another notable property is the inequality $\langle i_{\rm FB}^k \rangle \leq \langle i_{\rm FB}^{k^\prime} \rangle$ when $k < k^\prime$, which reflects that higher-order Markovian feedback is less restrictive.
We note that the information not fed back to system, $\langle i_{\rm QCT} \rangle - \langle i_{\rm FB}^{k} \rangle$ was introduced and named backward transfer entropy \cite{ito2016backward} in the classical regime.

Now, a refinement of the SL (\ref{eq:GSL}) under $k$-th-order Markovian feedback is obtained as  
\begin{equation}
    \langle \sigma \rangle \geq -\langle i_{\rm FB}^k \rangle. \label{eq:GSL_BQC} 
\end{equation}
This provides a tighter bound for the entropy reduction by accounting for the Markovianity of feedback, while the inequality (\ref{eq:GSL}) serves as the universal bound regardless of the feedback protocol details. 
The FT can also be generalized by incorporating $i_{\rm FB}^k$ as
\begin{equation}
    \langle e^{-\sigma-i_{\rm FB}^k} \rangle = 1,  \label{eq:GFT_BQC}
\end{equation}
which reveals the fundamental relationship between $\sigma$ and $i_{\rm FB}^k$ at the trajectory level.

\begin{figure}[]
\begin{center}
\includegraphics[width=0.45\textwidth]{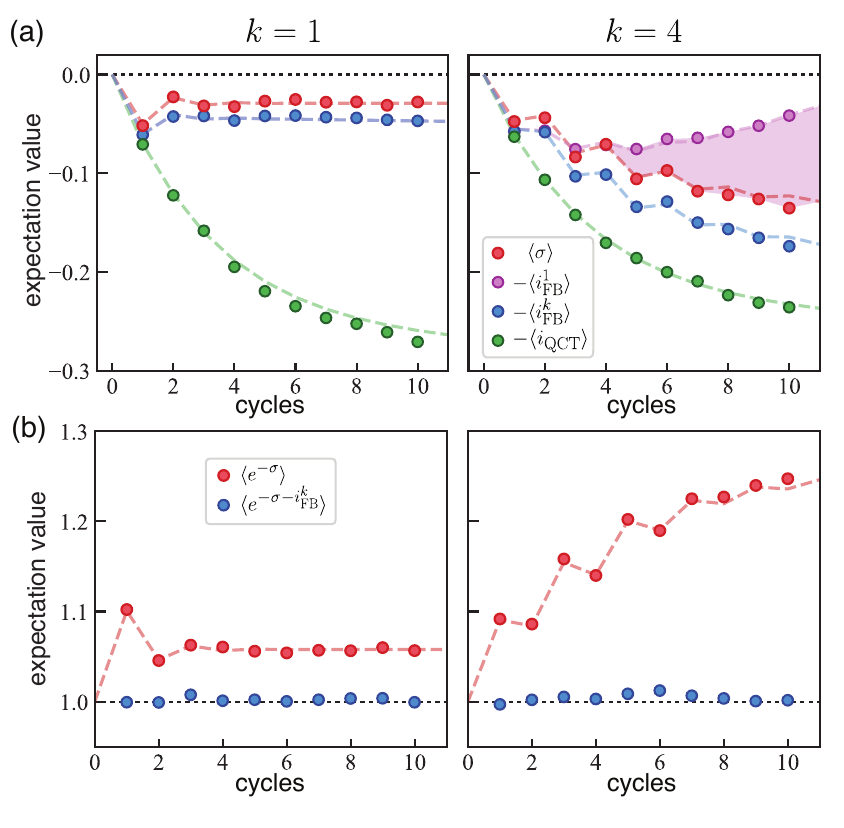}
\caption{Generalized second law and fluctuation theorem under $k$-th-order Markovian feedback. (a) Experimentally measured entropy production $\langle \sigma \rangle$, lower bound for the entropy production under Markovian feedback $-\langle i_{\rm FB}^{1} \rangle$, negative values of the net information fed back through $k$-th-order Markovian feedback $-\langle i_{\rm FB}^{k} \rangle$, and negative QC-transfer entropy $-\langle i_{\rm QCT} \rangle$ are plotted against the number of feedback cycles for $k=1,4$. The purple-shaded region indicates the entropy reduction beyond the Markovian bound $\langle \sigma \rangle < -\langle i_{\rm FB}^{1} \rangle$. (b) Experimentally measured $\langle e^{-\sigma}\rangle$ and $\langle e^{-\sigma -i_{\rm FB}^k}\rangle$ are plotted against the number of feedback cycles. In both (a) and (b), dotted lines represent theoretical predictions corresponding to the same colored plots. Error bars, which represent the standard error of the mean, are smaller than the marker sizes.}
\label{fig:nonMarkovExp_GSL}
\end{center}
\end{figure}

We apply this theoretical framework to a series of experiments where $k$-th-order Markovian feedback for $k=1,2,3,4$ is implemented.
The experiment still follows the general sequence shown in Fig.~\ref{fig:setup_and_sequence} but now with the feedback sequences as shown in Figs.~\ref{fig:nonMarkovExp}(a) and (b) for $k=1$ and $k=2$, respectively. Here, we perform stabilization of the excited state $\ket{1}$ instead of the ground state $\ket{0}$:  $\theta_n=\pi$ when all of the last $k$ outcomes are $0$ (i.e., $(y_{n-k+1},\dots,y_{n}) =(0,\dots,0)$), and $\theta_n=0$ otherwise. 
Measurement results before the last $k$ are erased and overwritten on the FPGA performing the feedback.
Note that at the beginning of an experimental trial, when the number of feedback cycles $n$ is smaller than the non-Markovianity $k$, we set $\theta_n = \pi$ if all of the last $n$ outcomes are $0$, and $\theta_n = 0$ otherwise.
Again, 10 successive cycles of measurement and feedback are applied for one experimental trial, which is repeated 10000 times.

The data in Fig.~\ref{fig:nonMarkovExp}(c) shows an improvement in the excited state stabilization with an increase in non-Markovianity, or memory length, $k$.  The more measurements are taken into consideration, the more accurately the system's state can be determined. The advantage of non-Markovian feedback can also be observed in terms of the information-thermodynamic efficiency $\eta$ \cite{toyabe2010experimental,koski2014experimental,masuyama2018information}, which quantifies the proportion of total information gain that can be converted to the entropy reduction. This efficiency is defined as 
\begin{equation}
    \label{eq:eff}
    \eta \equiv \frac{-\langle \sigma \rangle}{\langle i_{\rm QCT} \rangle}.
\end{equation}
As shown in Fig.~\ref{fig:nonMarkovExp}(d), $\eta$ has the same value for all $k$ when the number of feedback cycles $n$ is less than $k$, while it increases with $k$ for a larger number of cycles. 
However, we note that this efficiency does not continue to increase indefinitely with increasing $k$: the improvement from $k=3$ to $k=4$ is less significant than that from $k=1$ to $k=2$. A more detailed discussion of this behavior is provided in the Supplemental Material.

Fig.~\ref{fig:nonMarkovExp_GSL}(a) shows the entropy reduction $\langle \sigma \rangle<0$ as a result of the excited-state stabilization. $\langle \sigma \rangle$ is always lower-bounded by $-\langle i_{\rm FB}^k\rangle$ in accordance with (\ref{eq:GSL_BQC}).
However, it is not lower-bounded by the Markovian feedback information value $-\langle i_{\rm FB}^1 \rangle$ when the applied sequence is non-Markovian, e.g. with $k\geq 2$. 
This is highlighted with the shaded purple region in Fig.~\ref{fig:nonMarkovExp_GSL}(a), explicitly representing the thermodynamic advantage of non-Markovian feedback over Markovian feedback.
We note that the difference in $-\langle i_{\rm FB}^1\rangle$ between $k=1$ and $k=4$ in Fig. \ref{fig:nonMarkovExp_GSL}(a) is due to slight run-to-run variations in experimental parameters (see Supplemental Materials Sec. \ref{Ss:exp} for more details).
Fig.~\ref{fig:nonMarkovExp_GSL}(b) shows the generalized FT (\ref{eq:GFT_BQC}) incorporating feedback-induced information flow $i_{\rm FB}^k$. The modified FT is satisfied for all $k$ with $\langle e^{-\sigma -i_{\rm FB}^k}\rangle$ staying near unity throughout the experiment. This demonstrates the relevance of $i_{\rm FB}^k$ to thermodynamic benefit at a single trajectory level.

\section{Conclusion}
In this work, we experimentally verified the generalized second law of thermodynamics and fluctuation theorem, which incorporate measures of quantum information flow, using an SiV center coupled to a diamond nanocavity. Through a series of experiments varying the Markovianity and memory length of feedback protocols, we quantified the thermodynamic advantage of non-Markovian over Markovian feedback. To enable the analysis of the experimental protocols, we have extended the theoretical framework to describe the thermodynamic benefits resulting from the causal structure of the feedback. Our work directly probes the fundamental relationship between thermodynamic benefit and information flow in quantum systems, and explicitly reveals the thermodynamic advantage of non-Markovianity in thermodynamic terms. 

While we focus on a simple quantum system in this experiment, our work provides a framework that could be extended to more general and complex quantum systems. For example, it would be interesting to experimentally investigate finite temperature regime $\beta > 0$, where the energetic contribution also play a significant role. Studies in these systems could yield new insights for improved quantum control. The applications include efficient quantum feedback design for quantum state preparation and stabilization~\cite{sivak2022model, johnson2017} and quantum error correction \cite{li2021, googleqec2023}, which can be seen as processes controllably removing entropy from a quantum system. Furthermore, our work would pave the way for exploring the thermodynamic impact of real-time state estimation such as the quantum Kalman filter \cite{wiseman2009quantum}, which will play an essential role in achieving optimal control.

Since the thermodynamic information measures used in our work put strict bounds on the entropy production of a feedback and measurement process, they tell us the achievable state purity and fidelity. This paves the way for more optimal quantum measurement and feedback strategies. In particular, when detailed prior knowledge of the evolution and measurement implemented in the experiment is available, the information-thermodynamic efficiency can serve as a more appropriate benchmark for evaluating a feedback process than, for example, state fidelity or state purity. This is due to its ability to quantify how close the entropy reduction is to the information extracted from the system through the measurement instead of quantifying how close the system is to a desired target state. In other words, the information-thermodynamic efficiency is saturated when a feedback protocol completely uses the information that is extracted from the system through the measurement, whereas the state fidelity or purity is saturated only when the system is perfectly stabilized.
Furthermore, since in the classical regime measures of classical information flow such as the transfer entropy are relevant in feedback efficiency evaluation \cite{hartich2016sensory}, we expect their quantum counterparts covered in our work to play an important role in evaluating and improving quantum feedback protocols. 

Other quantum systems such as quantum heat engines and batteries, proposed quantum analogs of classical heat engines \cite{tajima2021superconducting,campaioli2017enhancing}, can also benefit from our nonequilibrium thermodynamic framework, including successive quantum feedback control, for example, for the robustness to external noise.

\begin{acknowledgments}
We acknowledge helpful discussions and technical support from Yan Qi Huan, Yan-Cheng Wei, Can Knaut, Madison Sutula, Gefen Baranes, and Daniel Assumpcao. We thank Jim MacArthur for electronics support. We also thank Ken Funo for the fruitful discussion. The device was fabricated at the Harvard Center for Nanoscale Systems, NSF award no. 2025158. 
T.Y. is supported by World-leading Innovative Graduate Study Program for Materials Research, Information, and Technology (MERIT-WINGS) of the University of Tokyo. T.Y. is also supported by JSPS KAKENHI Grant No. JP23KJ0672.
E.N.K. acknowledges support from an NSF GRFP No. DGE1745303.
N.Y. wishes to thank JST PRESTO No. JPMJPR2119, JST Grant Number JPMJPF2221, JST CREST Grant Number JPMJCR23I4, IBM Quantum, JST ASPIRE Grant Number JPMJAP2316.
T.S. is supported by JST ERATO-FS Grant Number JPMJER2204, JST ERATO Grant Number JPMJER2302, Japan, and JST CREST Grant Number JPMJCR20C1. N.Y. and T.S. are also supported by Institute of AI and Beyond of the University of Tokyo.
\end{acknowledgments}

\begin{appendix}

\section{Optical readout and $\delta_0$, $\delta_1$ tuning} \label{apps:opt_readout}
\begin{figure*}[]
\begin{center}
\includegraphics[width=0.75\textwidth]{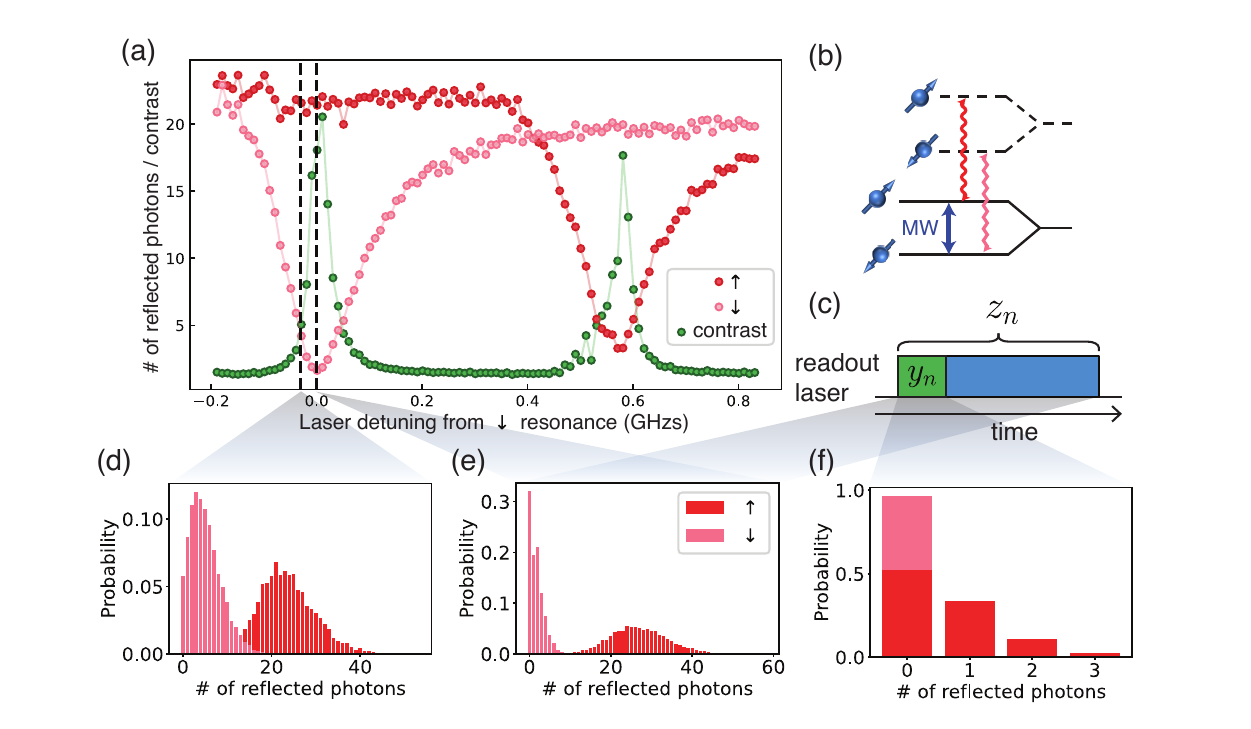}
\caption{Optical readout of the SiV electron spin state. (a) Spin state-dependent number of reflected photons and optical contrast in function of readout laser frequency detuning. (b) The energy level structure of the SiV, showing optical transitions in pink and red, and showing microwave-accessible transitions in dark blue. (c) Spin readout sequence, with the shorter readout ($y_n$) used for feedback and the longer readout ($z_n$) used for state tomography. Reflected photon count distribution for (d) off-resonant readout (corresponding to laser readout frequency of the left vertical dashed line in (a)), (e) on-resonant readout (corresponding to laser readout frequency of the right vertical dashed line in (a)), and (f) shorter readout.}
\label{fig:SI_readout}
\end{center}
\end{figure*}

The Purcell-enhanced frequency-dependent reflectivity of the SiV-cavity system exhibits large spin state-dependent reflection intensity contrast around the spin-like optical transitions. The contrast value is set by the SiV-cavity cooperativity, as well as the detuning of the SiV and cavity transitions. The cavity transition frequency is tuned so as to achieve the highest contrast when on resonance with the SiV transition. The SiV spin state is read out in the $\hat{\sigma}_z$-basis by shining light resonant with the spin-like transition for one of the spin states and monitoring the reflected photons. The resulting reflected photon distribution obeys Poissonian statistics with the average photon number dependent on the spin state. The photon distribution is then thresholded with a chosen value $n$ to determine the spin state (e.g., less than $n$ photons corresponds to $\ket{\downarrow}$, $n$ photons or more to $\ket{\uparrow}$. Here, up-spin and down-spin state correspond to the excited state $\ket{1}$ and ground state $\ket{0}$). Every readout consists first of a short readout ($y_n$), the result of which is stored by a fast FPGA for feed-forward decision making, and a longer readout with the result sent to a time tagger for data analysis ($z_n$).
The Kraus operators for this measurement become
\begin{equation}
\label{appeq:meas_op_cond}
    M_{y_n,z_n} \equiv 
    \begin{cases}
        \sqrt{1-\delta_0} \ket{0}\bra{0} \quad &(y_n=0,z_n=0) \\
        \sqrt{\delta_1} \ket{1}\bra{1} \quad &(y_n=0,z_n=1) \\
        \sqrt{\delta_0} \ket{0}\bra{0} \quad &(y_n=1,z_n=0) \\
        \sqrt{1-\delta_1} \ket{1}\bra{1} \quad &(y_n=1,z_n=1)
    \end{cases}
\end{equation}
The corresponding POVM operator for the outcome $y_n$ is defined as $E_{y_n} \equiv \sum_{z_n} M_{y_n,z_n}^\dag M_{y_n,z_n}$.

The strength of the read - and thus the values of $\delta_0$, $\delta_1$ - can be tuned in two ways. The first way is by changing the number of photons sent to the SiV-cavity system (Fig. \ref{fig:SI_readout}(e),(f)), with a higher number of incident photons resulting in an increase of the splitting between the Poisson distributions and thus a stronger read (the splitting increases linearly with average photon number, while the width of the distribution increases with the square root of the average photon number). The second way is by detuning the readout laser from the SiV resonance frequency (Fig. \ref{fig:SI_readout}(d),(e)), so that the optical contrast is reduced (see Fig. \ref{fig:SI_readout}) and therefore the splitting between the Poisson distribution is reduced. The parameters $\delta_0$, $\delta_1$ then obey the equations:
\begin{align*}
     \delta_0 = \sum_{i=n}^{\infty}\frac{\mu_{\downarrow}^ie^{-\mu_{\downarrow}}}{i!}\\
     \delta_1 = \sum_{i=0}^{n-1}\frac{\mu_{\uparrow}^ie^{-\mu_{\uparrow}}}{i!},
\end{align*}
where $\mu_{\downarrow/\uparrow}$ is the average reflected photon number for the $|\downarrow/\uparrow \rangle$ state, and $n$ is the threshold value to distinguish the SiV state based on the number of reflected photons. Experimentally, the parameters $\delta_0$, $\delta_1$ are explicitly calculated from the sampled data as
\begin{align*}
     \delta_0 &\equiv \frac{\mathrm{NUM}(y=1,z=0)}{\mathrm{NUM}(y=0,z=0)+\mathrm{NUM}(y=1,z=0)}, \\
     \delta_1 &\equiv \frac{\mathrm{NUM}(y=0,z=1)}{\mathrm{NUM}(y=0,z=1)+\mathrm{NUM}(y=1,z=1)},
\end{align*}
where $\mathrm{NUM}(y,z)$ represents the number that the measurement outcome set $(y,z)$ is sampled in the experiment.

\section{Thermodynamics under iterative quantum feedback}\label{apps:thermo_QCTE}
We review quantum thermodynamics under iterative measurement and feedback and detail the SL and FT generalized with the QC-transfer entropy, given by Eqs.~(\ref{eq:GSL}) and (\ref{eq:GFT}) in the main text.
While these generalized thermodynamic laws have already been introduced in Ref.~\cite{yada2022quantum}, we restate them here in a slightly more general setup, for the completeness of the paper.
We note that the generalized SL (\ref{eq:GSL_BQC}) and FT (\ref{eq:GFT_BQC}) incorporating $i_{\rm FB}^k$, which are newly derived in this paper, are discussed in Appendix~\ref{apps:Qthermo_BQC}.

\subsection{Setup}\label{appss:setup}

\begin{figure}[]
\begin{center}
\includegraphics[width=0.5\textwidth]{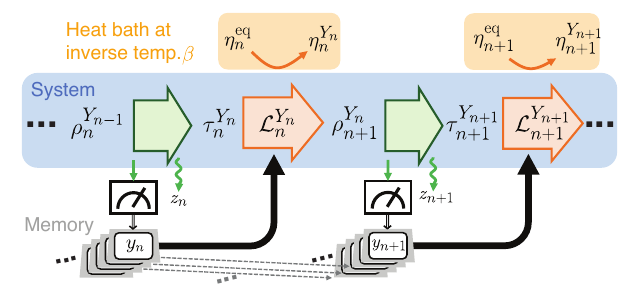}
\caption{Schematic of the quantum system under iterative measurement and feedback interacting with a heat bath. Here, $y_n$ denotes the $n$-th measurement outcome, and $z_n$ denotes the label that is not stored in memory nor utilized in feedback. The conditional density operators $\rho_n^{Y_{n-1}}$ and $\tau_n^{Y_{n}}$ denote the state before and after the $n$-th measurement, respectively, and $\mathcal{L}_n^{Y_n}$ represents the $n$-th feedback process. $\eta_n$ and $\eta_n^{Y_n}$ are states of the heat bath before and after the $n$-th feedback process, respectively.}
\label{fig:app_setup}
\end{center}
\end{figure}

In this work, we consider the quantum system under iterative measurement and feedback interacting with the heat bath at inverse temperature $\beta$, as shown in Fig.~\ref{fig:app_setup}. 
This includes not only the situation of sequential discrete measurement and feedback but also the continuous feedback situation where the dynamics is described with the stochastic master equation \cite{breuer2002theory,wiseman2009quantum}. 
We describe the outcome of $n$-th measurement as $y_n$, and all the outcomes until $y_n$ as $Y_n\equiv(y_1,\dots,y_n)$.
The total number of feedback cycle is denoted as $N$, and $\rho_{n}^{Y_{n-1}}$ and $\tau_n^{Y_n}$ represent the conditional density operators right before and after the $n$-th measurement process, respectively, for $1\leq n \leq N$.

The $n$-th measurement is described with the set of Kraus operators $\{M_{y_n,z_n}\}_{y_n,z_n}$, where the label $y_n$ is measured and utilized in the feedback while the label $z_n$ is lost. Therefore, conditional density operator after the $n$-th measurement becomes $P[y_n|Y_{n-1}] \tau_n^{Y_{n-1},y_n} = \sum_{z_n} M_{y_n,z_n} \rho_{n}^{Y_{n-1}} M_{y_n,z_n}^\dag$.
The $n$-th feedback process and heat-bath interaction are described as $\mathcal{L}_n^{Y_n} (\tau_n^{Y_n}) \equiv \tr_{\rm B}[U^{Y_n} (\tau_n^{Y_n}\otimes\eta^{\mathrm{eq}}_{n}) U^{Y_n\dag}]$, where $U^{Y_n}$ is the unitary operation to the composite system (quantum system and heat bath) depending on the measurement outcomes $Y_n$, and $\tr_{\rm B}[\cdot]$ denotes tracing out the heat-bath degree of freedom. 
$\eta^{\mathrm{eq}}_{n} \equiv \frac{e^{-\beta H_{\mathrm{B},n}}}{Z_{\beta}}$ denotes the thermal state of heat bath with Hamiltonian $H_{\mathrm{B},n}$ at inverse temperature $\beta$, where $Z_{\beta} \equiv \tr[e^{-\beta H_{\mathrm{B},n}}]$ is the partition function.
The set of Kraus operators for $\mathcal{L}_n^{Y_n}$, denoted as $\{L_{d_n}^{Y_n}\}_{d_n}$, is represented as $L_{(m,l)}^{Y_n} \equiv \bra{E_l}U^{Y_n}\ket{E_m} \sqrt{\frac{e^{-\beta E_m}}{Z_{\beta}}}$, where $\ket{E_l}$ and $\ket{E_m}$ are the energy eigenbases of the heat bath. The operator $L_{(m,l)}^{Y_n}$ represents the transition from $\ket{E_m}$ to $\ket{E_l}$, and the subscript $d_n$ represents to the labels $(m,l)$.
We note that Hamiltonian $H_{\mathrm{B},n}$, thermal state $\eta^{\mathrm{eq}}_{n}$, and energy eigenbasis $\{\ket{E_l}\}_l$ can also depend on $Y_n$, while its dependence is abbreviated here for simplicity.

Additionally, we employ a so-called two-time projective measurement scheme, where projective measurements (PMs) are performed before and after the $N$ feedback cycles. The bases for the initial and final PMs are denoted as $\{\ket{b_1}\}$ and $\{\ket{b_{N+1}}\}$.
In this case, the initial state $\rho_1$ must be diagonal in the basis $\{\ket{b_1}\}$, and the final state, which we denote by $\rho_{N+1}$, becomes diagonal in $\{\ket{b_{N+1}}\}$. The final state $\rho_{N+1}$ is defined as $\rho_{N+1}\equiv \sum_{Y_N} P[Y_N] \rho_{N+1}^{Y_N}$, where the conditional state $\rho_{N+1}^{Y_{N}}$ is described as $\rho_{N+1}^{Y_{N}} \equiv \sum_{b_{N+1}} \Pi_{b_{N+1}} \mathcal{L}_N^{Y_N} (\tau_N^{Y_N}) \Pi_{b_{N+1}}$ with the rank-one projector $\Pi_{b_{N+1}} \equiv \ket{b_{N+1}}\bra{b_{N+1}}$.
These PMs enable us to quantify the entropy change at the trajectory level, as discussed later in Appendix~\ref{apps:Qthermo_BQC}.
It is the standard scheme of quantum thermodynamics \cite{funo2018quantum,landi2021irreversible,manzano2022quantum}, and has also been employed in some experiments \cite{camati2016experimental,masuyama2018information,naghiloo2018information}.

Finally, we note that in a series of previous works~\cite{strasberg2017quantum,strasberg2019operational,strasberg2019stochastic,strasberg2019repeated,strasberg2020thermodynamics,strasberg2022quantum}, a thermodynamic framework for general stochastic quantum dynamics has been established, which includes the setup of iterative quantum feedback. However, these works differ from our framework significantly, both in their underlying conceptual approach and in the thermodynamic laws derived. Specifically, their framework aims at establishing thermodynamic consistency by explicitly including ancilla systems (such as memory) and considering thermodynamics at the level of composite system (system and ancilla). In contrast, our framework focuses on how measurement and feedback operations modify the thermodynamic behavior of the system itself. A more detailed discussion of these differences is provided in the Supplemental Material.

\subsection{Generalized second law} \label{appss:thermo_QCTE_GSL}
The average heat transferred from the heat bath until the $N$-th feedback process is defined as \cite{funo2018quantum,manzano2022quantum}
\begin{equation}
\label{eq:ens_Q}
    \langle Q \rangle \equiv \sum_{n=1}^{N} \sum_{Y_n} P[Y_n] \tr[H_{\mathrm{B},n}(\eta^{\mathrm{eq}}_{n}-\eta^{Y_n}_{n})],
\end{equation}
where $\eta^{Y_n}_{n}\equiv \tr_{\rm S}[U^{Y_n} (\tau_n^{Y_n} \otimes \eta^{\mathrm{eq}}_{n}) U^{Y_n\dag}]$ represents the state of $n$-th heat bath after $n$-th feedback process. We also define the heat income when the measurement outcomes are $Y_n$ as $Q_n^{Y_n} \equiv \tr[H_{\mathrm{B},n}(\eta^{\mathrm{eq}}_{n}-\eta^{Y_n}_{n})]$.
The averaged entropy production is defined as 
\begin{equation}
\label{eq:ens_ent_prd}
    \langle \sigma \rangle \equiv  S(\rho_{N+1}) - S(\rho_1) - \beta \langle Q\rangle,
\end{equation}
where $S(\rho) \equiv -\tr[\rho\ln\rho]$ denotes the von Neumann entropy. The entropy of the initial and final states $S(\rho_{1})$ and $S(\rho_{N+1})$ correspond to $\langle s_i \rangle $ and $\langle s_f \rangle$ in the main text.
The QC-transfer entropy, which is a measure of total information gain by iterative measurements, is defined as \cite{yada2022quantum}
\begin{equation}
    \label{eq:ens_QCTE}
    \langle i_{\mathrm{QCT}} \rangle = \sum_{n=1}^{N}  S(\rho_n|Y_{n-1}) -S(\tau_n|Y_{n}),
\end{equation}
where $S(\rho|y)\equiv \sum_y P[y]S(\rho^y)$ represents the conditional von Neumann entropy. Eq.~(\ref{eq:e_QCT}), which represents the increment of $ \langle i_{\mathrm{QCT}} \rangle$ by the $n$-th measurement, is consistent to the definition of Eq.~(\ref{eq:ens_QCTE}).
We will show later in Appendix~\ref{appss:TE}, that the QC-transfer entropy can be interpreted as the quantum counterpart of the (classical) transfer entropy \cite{schreiber2000measuring}, which is a widely-used measure of classical information flow.

With these thermodynamic quantities and information terms, we can derive the generalized SL~(\ref{eq:GSL}):
\begin{equation*}
    \langle \sigma \rangle \geq  -\langle i_{\rm QCT}\rangle.
\end{equation*}
The derivation of this inequality is already performed in Ref.~\cite{yada2022quantum}. Also, we will provide the derivation of the tighter second law (\ref{eq:GSL_BQC}) in Appendix~\ref{appss:GSL_BQC}.

\subsection{Generalized fluctuation theorem}\label{appss:thermo_QCTE_GFT}
\begin{figure}[]
\begin{center}
\includegraphics[width=0.48\textwidth]{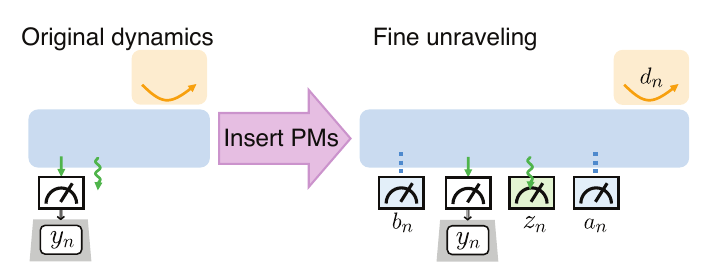}
\caption{Unraveling of the system dynamics. In the original dynamics at the ensemble level, only the measurement outcome $y_n$ is read out in the $n$-th cycle. On the other hand, under fine unraveling, the trajectories are designated not only by $y_n$ but also by the other labels $b_n, z_n, a_n,$ and $d_n$. Here, $b_n$ and $a_n$ denote the outcomes of the inserted PMs, which are performed on the diagonal bases of $\rho_{n}^{Y_{n-1}}$ and $\tau_n^{Y_n}$, respectively. The label $z_n$ denotes the outcome of the $n$-th measurement that is not stored in memory, and $d_n$ represents the label for the energy change of the heat bath.}
\label{fig:SI_unraveling}
\end{center}
\end{figure}

We introduce the definitions of the stochastic entropy production $\sigma$ and the stochastic QC-transfer entropy $i_{\rm QCT}$, with which the generalized FT $\langle e^{-\sigma - i_{\rm QCT}} \rangle =1$, given by Eq.~(\ref{eq:GFT_BQC}), is derived.
For that purpose, we need to properly decompose the averaged dynamics into trajectories by utilizing \textit{fine unraveling} introduced in Ref.~\cite{yada2022quantum}.

The essence of the fine unraveling is to insert the non-demolition projective measurements (PMs) before and after the measurement of $y_n$, in addition to the original iterative measurement and the heat-bath interaction, as shown in Fig.~\ref{fig:SI_unraveling}. 
The bases for these inserted PMs are determined by the spectral decompositions of the conditional density operators $\rho_n^{Y_{n-1}}$ and $\tau_n^{Y_n}$. Specifically, denoting their diagonalizations as
\begin{align*}
    \rho_{n}^{Y_{n-1}} &\equiv \sum_{b_{n}} p(b_{n}|Y_{n-1}) \ket{b_{n}}\bra{b_{n}}^{Y_{n-1}}, \\
    \tau_{n}^{Y_{n}} &\equiv \sum_{a_{n}} p(a_{n}|Y_n) \ket{a_{n}}\bra{a_{n}}^{Y_n},
\end{align*}
the bases for inserted PMs before and after the measurement of $y_n$ are given by the orthonormal bases $\{\ket{b_{n}}^{Y_{n-1}}\}$ and $\{\ket{a_{n}}^{Y_n}\}$, respectively.
The dependence of the diagonal bases $\ket{b_{n}}$ and $\ket{a_{n}}$ on the measurement outcomes $Y_{n-1}$ and $Y_n$ are abbreviated when it is clear from the context. 
It is noteworthy that their insertions do not destroy the original dynamics at the average level (i.e., the original dynamics is recovered by taking the ensemble average over the trajectory), since the bases of these PMs are diagonal with respect to the conditional density operators. We also note that the inserted PMs with the diagonal basis of $\rho_1$, $\{\ket{b_1}\}$, duplicates with the two-time PMs and thus can be abbreviated.

The fine-unraveled trajectory is designated with the outcomes of iterative measurements $\{y_n\}_{n=1}^N$ and $\{z_n\}_{n=1}^N$, heat-bath monitoring $\{d_n\}_{n=1}^N$, two-time PMs $b_1,b_{N+1}$, and the inserted PMs $\{b_n\}_{n=2}^{N}$ and $\{a_n\}_{n=1}^{N}$. 
We introduce the notation to denote all these labels together as 
\begin{equation*}
    \psi_N \equiv (\{b_n\}_{n=1}^{N+1}, \{y_n\}_{n=1}^N, \{z_n\}_{n=1}^N, \{a_n\}_{n=1}^{N}, \{d_n\}_{n=1}^N).
\end{equation*}
The set of Kraus operators for the fine unraveling is described as
\begin{equation}
    \left\{ \Pi_{b_{N+1}} \left(\prod_{n=1}^{N} L^{Y_n}_{d_n} \Pi_{a_{n}}^{Y_{n}} M_{y_n,z_n} \Pi_{b_{n}}^{Y_{n-1}} \right) \right\}_{\psi_N},
\end{equation}
where $\Pi_{b_{n}}^{Y_{n-1}}\equiv \ket{b_{n}}\bra{b_{n}}^{Y_{n-1}}$ and $\Pi_{a_{n}}^{Y_{n}}\equiv \ket{a_{n}}\bra{a_{n}}^{Y_{n}}$ for $1\leq n \leq N$ represent the rank-one projectors of inserted PMs, and $\Pi_{b_{N+1}} \equiv \ket{b_{N+1}}\bra{b_{N+1}}$ is the rank-one projector for the final PM. 
By utilizing this Kraus operator representation, the probability for the trajectory $\psi_N$ is calculated as 
\begin{equation}
\label{eq:prob_forward}
\begin{split}
    P_{\rm f}&[\psi_N] = \left| \bra{b_{N+1}} L_{d_N}^{Y_N} \Pi_{a_{N}}^{Y_{N}} M_{y_N,z_N} \ket{b_{N}}^{Y_{N-1}}\right|^2\\
    &\left\{\prod_{n=1}^{N-1} \left|{}^{Y_n}\bra{b_{n+1}} L_{d_n}^{Y_n} \Pi_{a_{n}}^{Y_{n}} M_{y_n,z_n} \ket{b_{n}}^{Y_{n-1}}\right|^2 \right\} p(b_1),
\end{split}
\end{equation}
where the subscript $\rm f$ of $P_{\rm f}[\psi_N]$ represents the fine unraveling.

Under the fine unraveling, both the stochastic entropy production $\sigma$ and the stochastic QC-transfer entropy $i_{\rm QCT}$ can be defined to each trajectory $\psi_N$.
The stochastic entropy production is defined as \cite{deffner2011nonequilibrium,horowitz2013entropy,hekking2013quantum}
\begin{equation}
    \sigma [\psi_N] \equiv -\ln p(b_{N+1}) + \ln p(b_1) - \beta Q[\psi_N],\label{eq:stoc_ent_prd}
\end{equation}
where the stochastic entropy of the final and initial states are $s_{f} \equiv -\ln p(b_{N+1})$ and $s_{i} \equiv -\ln p(b_1)$, respectively, and the stochastic heat input is defined as 
\begin{equation}
    Q[\psi_N] \equiv -\sum_{n=1}^{N} \Delta_{d_n}^{Y_n}.\label{eq:stoc_Q}
\end{equation}
Here, $\Delta_{(m,l)}^{Y_n} \equiv E_l^{Y_n}- E_m^{Y_n}$ represents the energy change of the heat bath due to the interaction with the system, under the condition that the past measurement outcomes are $Y_n$.
The stochastic QC-transfer entropy is defined as \cite{yada2022quantum}
\begin{equation}
    i_{\rm QCT} [\psi_N] \equiv \sum_{n=1}^{N} -\ln p(b_{n}|Y_{n-1}) + \ln p(a_{n}|Y_n),\label{eq:stoc_i_QCT}
\end{equation}
where $p(b_{n}|Y_{n-1})$ and $p(a_{n}|Y_n)$ are diagonal probability distribution for $\rho_n^{Y_{n-1}}$ and $\tau_n^{Y_n}$, respectively.
These stochastic quantities Eqs.~(\ref{eq:stoc_ent_prd}), (\ref{eq:stoc_Q}), and (\ref{eq:stoc_i_QCT})are reduced to Eqs.~(\ref{eq:ens_ent_prd}), (\ref{eq:ens_Q}), and (\ref{eq:ens_QCTE}) by taking the ensemble averages with respect to $\psi_N$, which means that the definitions of the averaged and stochastic quantities are consistent.

With these stochastic thermodynamic quantities and information term, we can derive the generalized FT~(\ref{eq:GFT}):
\begin{equation*}
    \langle e^{-\sigma-i_{\rm QCT}} \rangle =1.
\end{equation*}
The derivation of this equality is provided in Ref.~\cite{yada2022quantum}.

\section{Quantum thermodynamics depending on the causal structure of feedback} \label{apps:Qthermo_BQC}
We here introduce the measures of information transfer under $k$-th-order Markovian feedback, and elaborate on the generalized SL and FT, given by Eqs.~(\ref{eq:GSL_BQC}) and (\ref{eq:GFT_BQC}) in the main text.

\subsection{Information transfer under $k$-th-order Markovain feedback} \label{appss:BQC_k}
\begin{figure}[]
\begin{center}
\includegraphics[width=0.48\textwidth]{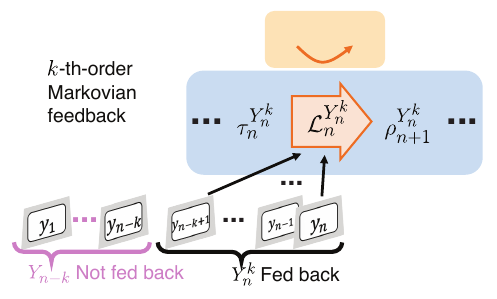}
\caption{Schematic of the $k$-th-order Markovian feedback setup. While the last $k$ measurement outcomes $Y_n^k$ are utilized to determine the feedback protocol $\mathcal{L}_n^{Y_n^k}$, it does not depend on the the earlier outcomes $Y_{n-k}$.}
\label{fig:k_th_order_Markov}
\end{center}
\end{figure}

To formulate the causal structure, we use the notion of $k$-th-order Markovian feedback, in which only the last $k$ measurement outcomes are utilized in feedback, as shown in Fig.~\ref{fig:k_th_order_Markov}.
We introduce the notation $Y_n^k \equiv (y_{n-k+1},\dots,y_{n-1},y_n)$ to describe the last $k$ measurement outcomes when $n\geq k$, while $Y_n^k$ is reduced to $Y_n$ when $n<k$.
The density operators conditioned only on the last $k$ outcomes are described as $\tau_n^{Y_n^k}$ and $\rho_{n+1}^{Y_{n}^k}$. These operators satisfy the relationships like $P[Y_n^k] \rho_{n+1}^{Y_n^k} \equiv \sum_{Y_{n-k}} P[Y_n] \rho_{n+1}^{Y_n}$, where $P[Y_n]$ and $P[Y_n^k]$ are the probabilities for the outcomes $Y_n$ and $Y_n^k$, respectively.
Under $k$-th-order Markovian feedback, the $n$-th feedback process and heat-bath interaction only depends on $Y_n^k$, and therefore the corresponding completely positive and trace-preserving map (CPTP map) can be described as $\mathcal{L}_n^{Y_n^k} (\tau_n^{Y_n^k}) \equiv \tr_{\rm B}[U^{Y_n^k} (\tau_n^{Y_n^k} \otimes\eta^{\mathrm{eq}}_{n}) U^{Y_n^k \dag}]$.

To derive the generalized SL under $k$-th-order Markovian feedback, we first introduce the net information that can be fed back to the system $\langle i_{\rm FB}^k \rangle$. It is defined as  
\begin{equation}
\label{eq:net_FB_info_k}
    \langle i_{\rm FB}^k \rangle \equiv \sum_{n=1}^{N}\left\{ \chi(\tau_n:Y_n^{k}) - \chi(\rho_{n+1}:Y_n^{k}) - \Delta S_{n}^{\rm BA}\right\},
\end{equation} 
where $\Delta S_{n}^{\rm BA}\equiv S(\tau_{n}) -S(\rho_{n})$ represents the entropy increase due to the back action of $n$-th measurement, and $\chi(\rho:y) \equiv S(\rho) -\sum_{y}p_y\rho^y$ denotes the Holevo information \cite{holevo1973bounds}, a well-known information measure to quantify the correlation between quantum system and classical variable. Considering the quantum-classical state $\sum_y P[y]\rho^y\otimes\ket{y}\bra{y}$ by explicitly introducing the classical degree of freedom $y$ as the register system with orthogonal basis $\{\ket{y}\}_y$, the Holevo information is interpreted as the quantum mutual information between quantum system and classical register. 
We can see from Eq.~(\ref{eq:net_FB_info_k}) that $\langle i_{\rm FB}^k \rangle$ quantifies the net information fed back, which is decomposed to two contributions: the correlation consumed during the $n$-th feedback process, $\chi(\tau_n:Y_n^{k}) - \chi(\rho_{n+1}:Y_n^{k})$, and the information lost due to the $n$-th measurement back action, $\Delta S_{n}^{\rm BA}$.

Under $k$-th-order Markovian feedback, $\langle i_{\rm FB}^k \rangle$ satisfies the inequality
\begin{equation}
\label{eq:BQCTE_size}
    \langle i_{\rm FB}^k \rangle  \leq \langle i_{\rm FB}^{k^\prime} \rangle  \leq \langle i_{\rm QCT} \rangle, \quad \mathrm{if}\ \  k^\prime > k.
\end{equation}
This inequality shows that the information fed back into the system is always smaller than the total information gain, and it decreases with the non-Markovianity of feedback $k$. 
The first inequality of Eq.~(\ref{eq:BQCTE_size}) is shown as 
\begin{widetext}
\begin{equation*}
    \langle i_{\rm FB}^{k^\prime} \rangle - \langle i_{\rm FB}^{k} \rangle= \sum_{n=1}^{N} \sum_{Y_n^{k^\prime}} P[Y_n^{k^\prime}] \left\{ S(\tau_n^{Y_n^{k^\prime}}\|\tau_n^{Y_n^{k}}) - S(\rho_{n+1}^{Y_{n}^{k^\prime}}\|\rho_{n+1}^{Y_n^{k}})\right\} \geq 0,
\end{equation*}
\end{widetext}
where $S(\rho\|\tau) \equiv \tr[\rho (\ln \rho - \ln \tau)]$ denotes the quantum relative entropy. We here utilize the relationships $\rho_{n+1}^{Y_{n}^{k^\prime}} = \mathcal{L}_n^{Y_n^k}(\tau_n^{Y_n^{k^\prime}})$ and $\rho_{n+1}^{Y_n^{k}} = \mathcal{L}_n^{Y_n^k}(\tau_n^{Y_n^{k}})$, and the monotonicity of relative entropy under CPTP map $\mathcal{L}_n^{Y_n^k}$ \cite{petz2003monotonicity}.
The second inequality in Eq.~(\ref{eq:BQCTE_size}) can also be derived as
\begin{widetext}
\begin{equation*}
    \langle i_{\rm QCT} \rangle - \langle i_{\rm FB}^{k^\prime} \rangle 
    = \sum_{n=1}^{N} \sum_{Y_n} P[Y_n] \left\{ S(\tau_n^{Y_n}\|\tau_n^{Y_n^{k^\prime}}) - S(\rho_{n+1}^{Y_n}\|\rho_{n+1}^{Y_n^{k^\prime}})\right\} + \chi(\rho_{N+1}:Y_N) \geq 0, 
\end{equation*}
\end{widetext}
where we use the relationships $\rho_{n+1}^{Y_n} = \mathcal{L}_n^{Y_n^{k}}(\tau_n^{Y_n})$ and $\rho_{n+1}^{Y_n^{k^\prime}} = \mathcal{L}_n^{Y_n^{k}}(\tau_n^{Y_n^{k^\prime}})$, monotonicity of relative entropy under CPTP map $\mathcal{L}_n^{Y_n^{k}}$ \cite{petz2003monotonicity}, and the positivity of Holevo information $\chi(\rho_{N+1}:Y_N) \geq 0$ for the derivation.

Here, we further introduce the information measure named backward QC-transfer entropy, which is wasted due to the restriction in feedback structure. As shown in Fig.~\ref{fig:k_th_order_Markov}, while a portion of information gain $\langle i_{\rm QCT} \rangle$ is fed back to system, another portion is wasted without being utilized in the feedback protocol. Under $k$-th-order Markovian feedback, it is defined as 
\begin{equation}
    \begin{split}
        I_{\mathrm{BQC}}^{k} &\equiv \langle i_{\rm QCT} \rangle - \langle i_{\rm FB}^k \rangle \\
        &= \left\{\sum_{n=k+1}^{N} \chi (\tau_{n}:Y_{n-k}|Y_n^k) - \chi (\rho_{n+1}:Y_{n-k}|Y_n^k)\right\} \\
        &   + \chi(\rho_{N+1}:Y_N).
    \end{split}\label{eq:BQCTE_def}
\end{equation}
The conditional Holevo information is defined as $\chi(\rho:y_2|y_1) \equiv \sum_{y_1} P[y_1]S(\rho^{y_1}) -\sum_{y_1,y_2} P[y_1,y_2]S(\rho^{y_1,y_2})$ for the ensemble of the quantum state $\rho^{y_1,y_2}$ with the probability $P[y_1,y_2]$. Here, $P[y_1] \equiv \sum_{y_2}P[y_1,y_2]$ is the marginal probability, $\rho^{y_1} \equiv \sum_{y_2}\frac{P[y_1,y_2]}{P[y_1]}\rho^{y_1,y_2}$ is the quantum state conditioned only on $y_1$, and $\rho \equiv \sum_{y_1,y_2} P[y_1,y_2]\rho^{y_1,y_2}$ is the averaged quantum state.
Considering the quantum-classical-classical state $\sum_{y_1,y_2} P[y_1,y_2]\rho^{y_1,y_2}\otimes\ket{y_1}\bra{y_1}\otimes\ket{y_2}\bra{y_2}$ with the classical registers with orthogonal basis $\{\ket{y_1}\}_{y_1}$ and $\{\ket{y_2}\}_{y_2}$, the conditional Holevo information is equivalent to the conditional quantum mutual information which quantifies the correlation between quantum system and $y_2$ under the condition of $y_1$.
Therefore, the term $\chi (\tau_{n}:Y_{n-k}|Y_n^k)$ quantifies the system-memory correlation which is not utilized in the feedback process from the $n$-th time onward, as shown in Fig.~\ref{fig:k_th_order_Markov}, and $\chi (\tau_{n}:Y_{n-k}|Y_n^k) - \chi (\rho_{n}:Y_{n-k}|Y_n^k)$ means the amount of unusable correlation that is lost during the feedback process $\mathcal{L}_n^{Y_n^k}$. 
Therefore, we can see from Eq.~(\ref{eq:BQCTE_def}) that $I_{\mathrm{BQC}}^k$ quantifies the wasted correlations between the system and memory as the summation of the two contributions: the correlations lost during the dynamics (i.e., the terms in curly brackets) and the correlations remained at the end (i.e., $\chi(\rho_{N+1}:Y_N)$).
We can also prove that when the dynamics is fully classical, $I_{\mathrm{BQC}}^k$ is reduced to the (classical) backward transfer entropy \cite{ito2016backward}, which is the measure of wasted classical correlation due to the restriction in the causal structure of feedback (see Appendix~\ref{appss:BTE} for the proof).

\subsection{Generalized second law under $k$-th-order Markovian feedback} \label{appss:GSL_BQC}
We derive the SL generalized with the net information fed back $\langle i_{\rm FB}^k\rangle$, described as
\begin{equation}
    \langle \sigma \rangle \geq -\langle i_{\rm FB}^k \rangle \geq -\langle i_{\rm QCT}\rangle. \label{appeq:GSL_BQC}
\end{equation}
The second inequality has already been shown in Appendix~\ref{appss:BQC_k} as Eq.~(\ref{eq:BQCTE_size}).
The first inequality of Eq.~(\ref{appeq:GSL_BQC}) can also be derived by utilizing the positivity of the relative entropy, as follows:
\begin{widetext}
\begin{align*}
    \langle \sigma \rangle + \langle i_{\rm FB}^k\rangle &= \sum_{n=1}^N \sum_{Y_n^k} P[Y_n^k] \left\{S(\rho_{n+1}^{Y_n^k})-S(\tau_n^{Y_n^k})+ \beta \tr[H_{\mathrm{B},n}(\eta^{Y_n^k}_{n}-\eta^{\mathrm{eq}}_{n})]\right\} \\
    &= \sum_{n=1}^N \sum_{Y_n^k} P[Y_n^k] S(U^{Y_n^k}(\tau_n^{Y_n^k}\otimes\eta^{\mathrm{eq}}_{n})U^{Y_n^k \dag}\| \rho_{n+1}^{Y_n^k} \otimes \eta^{\mathrm{eq}}_{n}) \geq 0,
\end{align*}
\end{widetext}
where $\eta^{Y_n^k}_{n} \equiv \tr_{\rm S}[U^{Y_n^k} (\tau_n^{Y_n^k} \otimes \eta^{\mathrm{eq}}_{n}) U^{Y_n^k\dag}]$ represents the state of $n$-th heat bath after the $n$-th feedback process under the condition that the last $k$ measurement outcomes are $Y_n^k$.

\subsection{Generalized fluctuation theorem under $k$-th-order Markovian feedback} \label{appss:GFT_BQC}
To derive the generalized FT given by Eq.~(\ref{eq:GFT_BQC}), we need to introduce the unraveling such that both the stochastic entropy production $\sigma$ and stochastic information flow $i_{\rm FB}^k$ are properly defined to each trajectory.
Since $i_{\rm FB}^k$ cannot be defined to the fine unraveled trajectories discussed in Appendix \ref{appss:thermo_QCTE_GFT}, we here introduce another unraveling named \textit{$k$-th-order coarse unraveling}, where $i_{\rm FB}^k$ is well-defined to each trajectory.

The essential idea of $k$-th-order coarse unraveling is similar to that of fine unraveling, where each trajectory is designated with the labels of iterative measurement, heat-bath interaction, and inserted PMs, as shown in Fig.~\ref{fig:SI_unraveling}.
The only difference from the fine unraveling is that the basis of inserted PMs are the diagonal basis of $\rho_{n}^{Y_{n-1}^k}$ and $\tau_n^{Y_n^k}$, not $\rho_{n}^{Y_{n-1}}$ and $\tau_n^{Y_n}$. The density operators $\rho_{n}^{Y_{n-1}^k}$ and $\tau_n^{Y_n^k}$ are diagonalized as 
\begin{align*}
    \rho_{n}^{Y_{n-1}^k} &\equiv \sum_{b_{n}} p(b_{n}|Y_{n-1}^k) \ket{b_{n}}\bra{b_{n}}^{Y_{n-1}^k}, \\
    \tau_{n}^{Y_{n}^k} &\equiv \sum_{a_{n}} p(a_{n}|Y_n^k) \ket{a_{n}}\bra{a_{n}}^{Y_{n}^k},
\end{align*}
where the dependence of the diagonal bases $\ket{b_{n}}$ and $\ket{a_{n}}$ on the measurement outcomes $Y_{n-1}^k$ and $Y_{n}^k$ is abbreviated when it is clear from the context. 
The set of Kraus operators for this unraveling is described as
\begin{equation}
    \left\{\Pi_{b_{N+1}} \left(\prod_{n=1}^{N} L^{Y_n^k}_{d_n} \Pi_{a_{n}}^{Y_{n}^k} M_{y_n,z_n} \Pi_{b_{n}}^{Y_{n-1}^k} \right) \right\}_{\psi_N},
\end{equation}
where $\Pi_{b_{n}}^{Y_{n-1}^k}\equiv \ket{b_{n}}\bra{b_{n}}^{Y_{n-1}^k}$ and $\Pi_{a_{n}}^{Y_{n}^k} \equiv \ket{a_{n}}\bra{a_{n}}^{Y_{n}^k}$ denote the projectors of inserted PMs.
Therefore, the probability for the trajectory $\psi_N$ under $k$-th-order coarse unraveling can be calculated as 
\begin{equation}
\label{eq:prob_kMarkov}
\begin{split}
    P_{k,\rm c}&[\psi_N] = \left|\bra{b_{N+1}} L_{d_N}^{Y_N^k} \Pi_{a_N}^{Y_N^k} M_{y_N,z_N} \ket{b_{N}}^{Y_{N-1}^k}\right|^2 \\
    &\left\{\prod_{n=1}^{N-1} \left|{}^{Y_n^k}\bra{b_{n+1}} L_{d_n}^{Y_n^k} \Pi_{a_n}^{Y_n^k} M_{y_n,z_n} \ket{b_{n}}^{Y_{n-1}^k}\right|^2 \right\} p(b_1).
\end{split}
\end{equation}
Due to the choice of the PM bases, we can recover the original dynamics of density operators $\rho_{n}^{Y_{n-1}^k}$ and $\tau_{n}^{Y_{n}^k}$ by taking the ensemble average, if the feedback protocol is $k$-th-order Markovian.
We note that as can be seen from Eqs.~(\ref{eq:prob_forward}) and (\ref{eq:prob_kMarkov}), the probability for the trajectories with the same label $\psi_N$ is different depending on the types of the unraveling.

Under $k$-th-order coarse unraveling, the stochastic entropy production can be defined in the same way as the fine unraveling, as described in Eq.~(\ref{eq:stoc_ent_prd}).
Furthermore, the stochastic information transfer $i_{\rm FB}^k$ is also well-defined:
\begin{multline}
\label{eq:stoc_net_FB_info_k}
    i_{\rm FB}^k [\psi_N] \equiv \left\{\sum_{n=1}^N \ln p (a_n|Y_n^k) -\ln p (b_{n+1}|Y_{n}^k) \right\} \\
    +\ln p(b_{N+1}) - \ln p (b_1).
\end{multline}
We can confirm that this definition of $i_{\rm FB}^k$ is consistent to the definition of $\langle i_{\rm FB}^k \rangle$ given by Eq.~(\ref{eq:net_FB_info_k}), by taking the ensemble average.
With these stochastic quantities, we can derive the generalized FT described as
\begin{equation*}
    \langle e^{-\sigma-i_{\rm FB}^k} \rangle = 1. \tag{\ref{eq:GFT_BQC}}
\end{equation*}
The derivation of this equality is basically the same as that of Eq.~(\ref{eq:GFT}) discussed in Ref.~\cite{yada2022quantum}. 
The detailed derivation of Eq.~(\ref{eq:GFT_BQC}) is provided in the Supplemental Material.

\section{Relationship between quantum and classical information measures}
\label{Appendix:quantum_classical}
When the dynamics are fully classical (i.e., can be described with a classical stochastic process), the quantum information measures are reduced to the corresponding classical measures.
Specifically, we show that the QC-transfer entropy $\langle i_{\rm QCT}\rangle$ and the backward QC-transfer entropy $I_{\rm BQC}^k \equiv \langle i_{\rm QCT}\rangle -\langle i_{\rm FB}^k\rangle$ are reduced to transfer entropy $\langle i_{\rm TE} \rangle$ \cite{schreiber2000measuring} and backward transfer entropy $I_{\rm BTE}^k$~\cite{ito2016backward}, respectively.

\subsection{Fully classical dynamics} \label{appss:fully_classical}
\begin{figure}[]
\begin{center}
\includegraphics[width=0.48\textwidth]{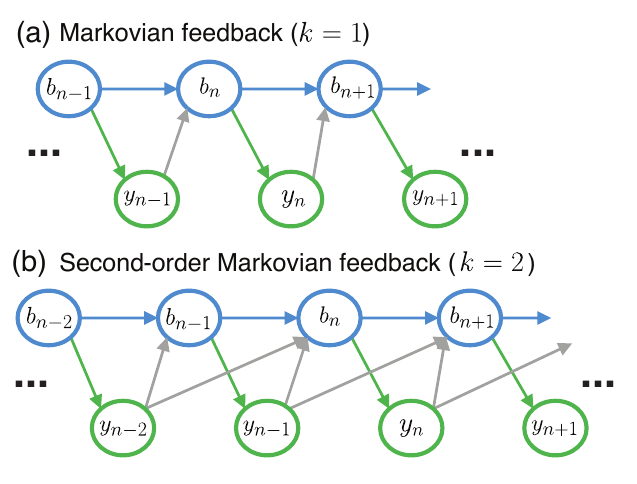}
\caption{Causal network representation of the classical stochastic processes under (a) Markovian and (b) second-order Markovian feedback. These fully classical dynamics are considered as special cases of our setup of quantum iterative feedback. Here, $\{b_n\}_{n=1}^N$ are the set of variables representing the system's dynamics, and $\{y_n\}_{n=1}^N$ denotes the set of variables for the measurement outcomes.}
\label{fig:causal}
\end{center}
\end{figure}

Our setup of iterative feedback shown in Fig.~\ref{fig:app_setup} generally exhibits quantum nature, such as the measurement back action. 
We here consider the special case of this setup where the dynamics are effectively reduced to the classical stochastic process.
Specifically, we consider the situation where the measurement operators are describe as 
\begin{equation}
    M_{y_{n},z_{n}} = \sum_{i} \sqrt{q(y_{n},z_{n}|i)} \ket{i}\bra{i}, \label{eq:classical_meas_op}
\end{equation}
and the feedback process satisfies 
\begin{equation}
    \mathcal{L}_n^{Y_n^k} (\ket{i}\bra{i}) = \sum_{j} w(j |i,Y_n^k) \ket{j}\bra{j}, \label{eq:classical_FB}
\end{equation}
where $\{\ket{i}\}_{i}$ is a set of fixed orthogonal basis.
Under these conditions, the system is always described as a classical mixture of $\{\ket{i}\}$ (i.e., both $\{\ket{a_n}^{Y_n^k}\}$ and $\{\ket{b_n}^{Y_{n-1}^k}\}$ are reduced to $\{\ket{i}\}$), and there is no measurement back action even at the trajectory level (i.e., $a_n =b_n$ for all $n$). 
The stochastic process of $\{b_n\}_{n=1}^{N}$ and $\{y_n\}_{n=1}^N$ in such fully classical situation is described in Fig.~\ref{fig:causal} with causal networks (see e.g., Ref.~\cite{jensen2007bayesian} for the review of causal network representation).

In such case, conditional density operators $\rho_n^{Y_{n-1}}$ and $\tau_n^{Y_{n}}$ can be represented as
\begin{align*}
\rho_n^{Y_{n-1}} &= \sum_{i} p(b_n = i|Y_{n-1}) \ket{i}\bra{i}, \\
\tau_n^{Y_{n}} &= \sum_{i} p(a_n = i|Y_{n}) \ket{i}\bra{i}.
\end{align*}
Since $\sum_{z_n} M_{y_n,z_n}\rho_n^{Y_{n-1}} M_{y_n,z_n}^\dag =P[y_n|Y_{n-1}] \tau_n^{Y_{n}}$ holds, we have the equality
\begin{equation}
\label{appeq:classical_markov}
    p(a_n = i|Y_{n}) = \frac{\sum_{z_n} q(y_{n},z_{n}|i) p(b_n = i|Y_{n-1}) }{P[y_n|Y_{n-1}]},
\end{equation}
which relates the diagonal probability distributions.

\subsection{Transfer entropy} \label{appss:TE}
In the setup of iterative quantum feedback, (classical) transfer entropy is defined as follows:
\begin{align}
    i_{\rm TE} &=\sum_{n=1}^{N} -\ln p(b_{n}|Y_{n-1}) +\ln p(b_{n}|Y_{n}), \label{eq:CTE_traj} \\
    \langle i_{\rm TE}\rangle &= \sum_{n=1}^{N} I(b_n:y_{n}|Y_{n-1}). \label{eq:CTE_ens}
\end{align}
Here, conditional probability $p(b_{n}|Y_{n})$ is defined as 
\begin{equation}
\label{appeq:cond_prob_b}
    p(b_{n}|Y_{n})\equiv \frac{p(y_{n}|b_n,Y_{n-1}) p(b_n|Y_{n-1})}{P[y_{n}|Y_{n-1}]},
\end{equation}
with $p(y_{n}|b_n,Y_{n-1}) \equiv \sum_{z_{n}} \| M_{y_{n},z_{n}}\ket{b_n}^{Y_{n-1}}\|^2$ and $P[y_{n}|Y_{n-1}] \equiv \sum_{z_{n}}\tr[M_{y_{n},z_{n}} \rho_n^{Y_{n-1}} M_{y_{n},z_{n}}^\dag]$. The conditional mutual information $I(b_n:y_{n}|Y_{n-1})$ quantifies the correlation newly generated between the system state $\ket{b_n}^{Y_{n-1}}$ and the memory, in the $n$-th measurement. 

When the dynamics is fully classical, we can show $i_{\rm TE}$ is reduced to the QC-transfer entropy, defined as 
\begin{equation*}
    i_{\rm QCT} \equiv \sum_{n=1}^{N} -\ln p(b_{n}|Y_{n-1}) + \ln p(a_{n}|Y_n).
\end{equation*}
The reduction follows from the fact that $p(a_n|Y_n) = p(b_n|Y_n)$, which implies the absence of measurement backaction.  
This equality can be derived from Eqs.~\eqref{appeq:classical_markov} and \eqref{appeq:cond_prob_b}.  
This reduction implies that if $i_{\rm QCT}$ and $i_{\rm TE}$ take different values, the system dynamics are not fully classical.

We now illustrate the difference between the averaged transfer entropy and the QC-transfer entropy through two simple examples that highlight the role of measurement backaction.  
To this end, we consider a two-level system and focus on the first measurement ($n=1$).  
We adopt the measurement operators used in our experiment [Eq.~\eqref{appeq:meas_op_cond}], and examine the following two cases:  
(i) $\delta_0 = \delta_1 = 0$ (i.e., projective measurement in the computational basis $\{\ket{0}, \ket{1}\}$), and  
(ii) $\delta_0 = \delta_1 = \frac{1}{2}$ (i.e., dephasing in the same basis).
The initial state $\rho_1$ is expressed via its spectral decomposition as  
$\rho_1 = \sum_{b_1 = 0,1} q(b_1) \ket{b_1} \bra{b_1}$.  
We define the overlaps between the eigenbasis $\{\ket{b_1=0}, \ket{b_1=1}\}$ of $\rho_1$ and the computational basis as  
$|\langle 0 | b_1=0 \rangle|^2 = |\langle 1 | b_1=1 \rangle|^2 = \alpha$, with $0 < \alpha < 1$.  
This setup ensures that the two basis sets are not identical.
In this setting, we compare the quantum and classical information measures:
\begin{align*}
\langle i_{\rm QCT} \rangle &=  S(\rho_1) - \sum_{y_1} P(y_1) S(\tau_1^{y_1}), \\
\langle i_{\rm TE} \rangle &=  I(b_1 : y_1).
\end{align*}

First, in case (i) with $\delta_0 = \delta_1 = 0$, the QC-transfer entropy becomes $\langle i_{\rm QCT} \rangle = S(\rho_1)$, since each conditional state $\tau_1^{y_1 = 0,1}$ becomes a pure state.  
The classical transfer entropy in this case is given by $\langle i_{\rm TE} \rangle = S(\rho_1) - h(\alpha)$, where $h(p) = -p \ln p - (1 - p) \ln (1 - p)$ is the binary entropy function.  
This follows from the joint probability $p(b_1 = i, y_1) = q(b_1 = i) |\langle b_1 = i | y_1 \rangle|^2$.  
Thus, we obtain the inequality $\langle i_{\rm QCT} \rangle > \langle i_{\rm TE} \rangle$.
In case (ii) with $\delta_0 = \delta_1 = \frac{1}{2}$, the classical transfer entropy becomes $\langle i_{\rm TE} \rangle = 0$, since the joint probability becomes $p(b_1, y_1) = \frac{1}{2} q(b_1)$, indicating statistical independence of $b_1$ and $y_1$.  
The QC-transfer entropy is given by $\langle i_{\rm QCT} \rangle = S(\rho_1) - S(\mathcal{D}(\rho_1))$, where $\mathcal{D}(\rho) = \ket{0}\bra{0} \rho \ket{0}\bra{0} + \ket{1}\bra{1} \rho \ket{1}\bra{1}$ represents the dephasing channel.  
If $q(0) \neq q(1)$, then $S(\mathcal{D}(\rho_1)) > S(\rho_1)$, implying $\langle i_{\rm QCT} \rangle < 0$, which leads to $\langle i_{\rm QCT} \rangle < \langle i_{\rm TE} \rangle$.
These two examples demonstrate that the QC-transfer entropy deviates from the classical transfer entropy due to measurement backaction, and that there is no fixed ordering between them in general.

\subsection{Backward transfer entropy} \label{appss:BTE}
We show that the backward QC-transfer entropy $I_{\rm BQC}^k$, introduced in Eq.~(\ref{eq:BQCTE_def}), is reduced to the (classical) backward transfer entropy when the dynamics is fully classical. 
Under the classical dynamics, $I_{\rm BQC}^k$ is represented as
\begin{align*}
    I_{\rm BQC}^k &= \left\{ \sum_{n=k+1}^{N} \chi (\tau_{n}:Y_{n-k}|Y_n^k) - \chi (\rho_{n+1}:Y_{n-k}|Y_n^k)\right\} \\
    &+ \chi(\rho_{N+1}:Y_N).\\
    &= \left\{ \sum_{n=k+1}^N I(b_{n}:Y_{n-k}|Y_n^k) - I(b_{n+1}:Y_{n-k}|Y_n^k) \right\} \\
    &+ I(b_{N+1}:Y_N), \\
\end{align*}
where the Holevo information is reduced to the classical mutual information.
In the following, we show that in fully classical dynamics under $k$-th-order Markovian feedback, $I_{\rm BQC}^k$ is equivalent to the backward transfer entropy $I_{\rm BTE}^k$, defined as \cite{ito2016backward,ito2016information}
\begin{equation}
    I_{\rm BTE}^k \equiv \sum_{n=k+1}^N I(b_{n}:y_{n-k}|Y_N^{N-(n-k)}) + I(b_{N+1}:Y_N^k).
\end{equation}

The equality $I_{\rm BQC}^k = I_{\rm BTE}^k$ can be derived as follows:
\begin{align*}
    I_{\rm BQC}^k &= \left\{ \sum_{n=k+1}^N I(b_{n}:Y_{n-k}|Y_n^k) - I(b_{n}:Y_{n-k-1}|Y_{n-1}^k) \right \} \\
    &+I(b_{N+1}:Y_N^k) \\
     &= \left \langle \ln \prod_{n=k+1}^N \frac{p(Y_{n-k}|b_{n},Y_n^k)}{P[Y_{n-k}|Y_n^k]} \frac{P[Y_{n-k-1}|Y_{n-1}^k]}{p(Y_{n-k-1}|b_{n},Y_{n-1}^k)}\right \rangle \\ &+I(b_{N+1}:Y_N^k) \\
     &= \left \langle \ln \prod_{n=k+1}^N \frac{p(Y_{n-k}|b_{n},Y_n^k)}{p(Y_{n-k-1}|b_{n},Y_{n}^{k+1})} \frac{P[Y_{n-k-1}|Y_{n-1}^k]}{P[Y_{n-k}|Y_n^k]}\right \rangle \\ &+I(b_{N+1}:Y_N^k) \\
     &= \left \langle \ln \left(\prod_{n=k+1}^N \frac{p(b_{n},Y_n^{k+1})}{p(b_{n},Y_{n}^{k})}\right) \frac{P[Y_N^k]}{P[Y_N]}\right \rangle +I(b_{N+1}:Y_N^k) \\
     &= \left \langle \ln \prod_{n=k+1}^N \frac{p(y_{n-k}|b_{n},Y_n^{k})}{P[y_{n-k}|Y_N^{N-(n-k)}]}\right \rangle +I(b_{N+1}:Y_N^k) \\
     &= \left \langle \ln \prod_{n=k+1}^N \frac{p(y_{n-k}|b_{n},Y_N^{N-(n-k)})}{P[y_{n-k}|Y_N^{N-(n-k)}]}\right \rangle +I(b_{N+1}:Y_N^k) \\
     &= \sum_{n=k+1}^N I(b_{n}:y_{n-k}|Y_N^{N-(n-k)}) +I(b_{N+1}:Y_N^k) \\
     &= I_{\rm BTE}^k, 
\end{align*}
where $\langle \cdot \rangle$ represents the ensemble average. In the derivation of the third equality, we utilize the relationship $p(Y_{n-k-1}|b_{n},Y_{n}^{k+1}) = p(Y_{n-k-1}|b_{n},Y_{n-1}^{k})$, which is equivalent to $p(y_n|b_n,Y_{n-1}) = p(y_n|b_n,Y_{n-1}^k)$. This equality follows from the chain rule $Y_{n-k-1}\rightarrow (b_n,Y_{n-1}^k) \rightarrow y_n$. In the derivation of the sixth equality, we use the relationship $p(y_{n-k}|b_{n},Y_N^{N-(n-k)}) = p(y_{n-k}|b_{n},Y_n^{k})$, which follows from the chain rule $y_{n-k}\rightarrow (b_n,Y_{n}^k) \rightarrow Y_N^{N-n}$ representing the $k$-th-order Markovianity of feedback. 
We can check these chain rules with the causal networks shown in Fig.~\ref{fig:causal}.

\end{appendix}

\clearpage
\bibliography{bibliography}



\onecolumngrid

\clearpage
\begin{center}
	\Large
	\textbf{Supplemental Material for ``Experimentally probing entropy reduction via iterative quantum information transfer"}
\end{center}

\setcounter{section}{0}
\setcounter{equation}{0}
\setcounter{figure}{0}
\setcounter{table}{0}
\setcounter{page}{1}
\renewcommand{\thesection}{S\arabic{section}}
\renewcommand{\theequation}{\thesection.\arabic{equation}}
\renewcommand{\thefigure}{S\arabic{figure}}
\renewcommand{\thetable}{S\arabic{table}}

\section{Experimental details} \label{Ss:exp}

\subsection{Experimental sequence}\label{Sss:exp_overveiw}

\begin{figure*}[]
\begin{center}
\includegraphics[width=1\textwidth]{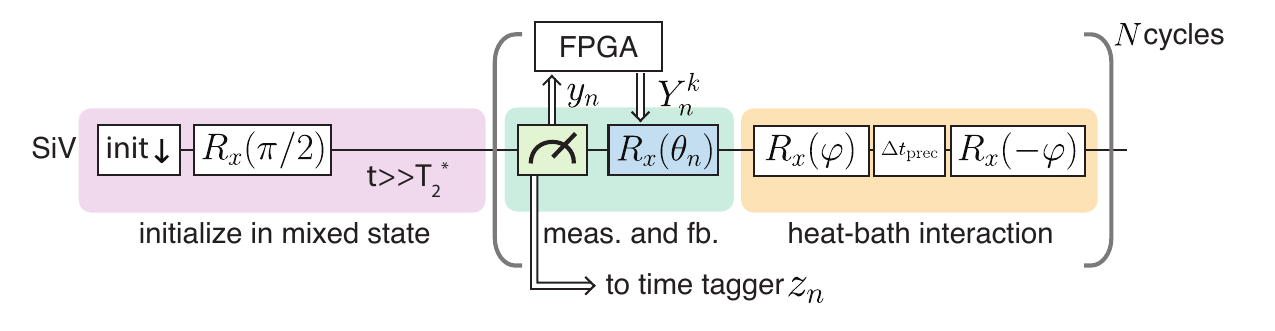}
\caption{Diagram for the experimental sequence. MW gates and measurements applied during different sections of the sequence are shown. In the preparation of the initial state, we prepare a maximally mixed state $\frac{1}{2} (\ket{0}\bra{0} + \ket{1}\bra{1})$ by applying the unitary gate $\hat{R}_x(\frac{\pi}{2})$ to $\ket{0}$ and dephasing with respect to the basis $\{\ket{0},\ket{1}\}$. In the measurement and feedback section, we perform measurements whose Kraus operators are represented by Eq.~(\ref{eq:SI_meas_op}) and apply a unitary gate depending on the past measurement outcomes $Y_n^k$. In the heat-bath interaction section, we perform unitary rotations $\hat{R}_x(\varphi)$ and $\hat{R}_x(-\varphi)$ before and after the precession for $\Delta t_{\rm prec}$ respectively.}
\label{fig:SI_experimental_sequence}
\end{center}
\end{figure*}

Each sequence in this work starts by preparing the SiV in a mixed state. This is achieved by first initializing the SiV in the $\ket{\downarrow}$ state, then rotating the SiV onto the equator with a $\hat{R}_x(\pi/2)$ pulse, and finally waiting for a time much longer than the SiV coherence time $T_2^*$ to ensure the SiV has decohered. We then follow with repetitions of manipulation blocks consisting first of a spin readout. A variable-angle rotation $\hat{R}_x(\theta_n)$ is then applied through microwave (MW) driving, with $\theta_n$ decided by the FPGA based on the previous readout outcome (in the Markovian case, $Y_n^1$) or the previous $k$ readout outcomes (in the non-Markovian case, $Y_n^k$). Finally, we implement a free precession and heat-batch interaction step. The SiV is first rotated by an angle $\varphi$, after which it both precesses by an angle $\phi$ and partially dephases through heat-bath interaction, after which we rotate the spin by an angle $-\varphi$.The experimental sequence is shown in detail in Fig. \ref{fig:SI_experimental_sequence}. We sample the experimental trajectories by repeating the experimental sequence multiple times (7000 times for the ground-state stabilization experiment and 10000 times for the excited-state stabilization experiment). 
In the following, we describe the experimental sequence with Kraus operators and quantum channels corresponding to the theoretical framework reviewed in Appendices \ref{apps:thermo_QCTE} and \ref{apps:Qthermo_BQC}.

The measurement operators of optical readout can be described as follows:
\begin{equation}
\label{eq:SI_meas_op}
    M_{y_n,z_n} \equiv 
    \begin{cases}
        \sqrt{1-\delta_0} \ket{0}\bra{0} \quad &(y_n=0,z_n=0) \\
        \sqrt{\delta_1} \ket{1}\bra{1} \quad &(y_n=0,z_n=1) \\
        \sqrt{\delta_0} \ket{0}\bra{0} \quad &(y_n=1,z_n=0) \\
        \sqrt{1-\delta_1} \ket{1}\bra{1} \quad &(y_n=1,z_n=1)
    \end{cases}
\end{equation}
The parameters $(\delta_0,\delta_1)$ are set as $(\delta_0,\delta_1)=(0.089,0.430)$ for the Markovian feedback experiment in Fig.~\ref{fig:MarkovExp}. 
In the series of experiments in Fig.~\ref{fig:nonMarkovExp_GSL}, $\delta_0,\delta_1$ are maintained almost constant for different values of $k$, although there are inevitable variations due to fluctuations in laser frequency and power; the parameters are set as follows: $(\delta_0,\delta_1) = (0.151,0.505)$ for $k=1$, $(\delta_0,\delta_1) = (0.157,0.513)$ for $k=2$, $(\delta_0,\delta_1) = (0.155,0.518)$ for $k=3$, and $(\delta_0,\delta_1) = (0.162,0.516)$ for $k=4$.
The details of this measurement process are discussed in Appendix~\ref{apps:opt_readout}.

The angle $\theta_n$ of the feedback pulse is determined in different ways in the ground state and excited state stabilization experiment (where we perform Markovian and non-Markovian feedback, respectively).
For the Markovian sequences depicted in Fig.~\ref{fig:setup_and_sequence}(d) in the main text, $\theta_n$ is chosen as follows:
\begin{align*}
    \theta_n = \pi, \,\,\, \mathrm{if} \,\, y_n = 1 \\
    \theta_n = 0,\,\,\, \mathrm{if} \,\, y_n = 0 .
\end{align*}
The non-Markovian sequences with memory length $k$ are designed as follows, as shown in Fig.~\ref{fig:nonMarkovExp}(a) and (b) in the main text:
\begin{align*}
    \theta_n = \pi, \,\,\, &\mathrm{if} \,\, \sum_{i=\mathrm{max}(0, n-k)}^n y_i = 0 \\
    &\mathrm{else:} \,\,\, \theta_n = 0 .
\end{align*}

During the heat-bath interaction process, unitary rotation and dephasing occur simultaneously.
This process is described with the quantum channel $\mathcal{N} \equiv \mathcal{U}_{-\varphi}^x \circ \mathcal{U}_{\phi}^z \circ \mathcal{D} \circ \mathcal{U}_{\varphi}^x$, where $x$-axis rotation channel $\mathcal{U}_{\varphi}^x$, precession channel $\mathcal{U}_{\phi}^z$, and dephasing channel $\mathcal{D}$ are defined as follows:
\begin{gather}
    \mathcal{U}_{\varphi}^x(\rho)  \equiv \hat{R}_x(\varphi) \rho \hat{R}_x(\varphi)^\dag, \\
    \mathcal{U}_{\phi}^z(\rho)  \equiv \hat{R}_z(\phi) \rho \hat{R}_z(\phi)^\dag, \\
    \mathcal{D}(\rho) \equiv  e^{-\alpha} \rho+ (1-e^{-\alpha}) \left(\ket{0}\bra{0}\rho\ket{0}\bra{0} +\ket{1}\bra{1}\rho\ket{1}\bra{1}\right),
\end{gather}
where $\alpha$ is the dephasing parameter. 
We can see that this process $\mathcal{N}$ can be interpreted as the interaction with the heat bath at inverse temperature $\beta = 0$ since the maximally mixed state is the fixed point of this channel.
In this case, the composite quantum map of the feedback and heat-bath interaction processes becomes $\mathcal{L}_n^{Y_n^k} = \mathcal{N} \circ \mathcal{U}_{\theta_n}^x$, where the dependence of $\theta_n$ on $Y_n^k$ is abbreviated here.
In the experiment of Fig.~\ref{fig:MarkovExp}, the parameters $(\varphi,\phi,\alpha)$ are determined as $(\varphi,\phi,\alpha) = (0.133 \pi, 0.280 \pi, 1.06 \times 10^{-2})$.
In the series of experiments in Fig.~\ref{fig:nonMarkovExp_GSL}, these parameters become $(\varphi,\phi,\alpha) = (0.151 \pi, 0.269 \pi, 1.02 \times 10^{-2})$. 
The implementation of these operations and the determination of the parameters are detailed in Sec.~\ref{Sss:gate_parameter}.

Finally, two-time PMs before and after the entire dynamics are both performed in $\ket{0},\ket{1}$ basis.
In fact, these PMs are already incorporated in the sequence shown in Fig.~\ref{fig:SI_experimental_sequence}, and therefore, there is no need to perform further measurements.
The initial PM is integrated in the first measurement with the operators $\{M_{y_1,z_1}\}_{y_1,z_1}$, as clear from their definitions in Eq.~(\ref{eq:SI_meas_op}) (the outcome of the initial PM is $b_1 = z_1$).
The final PM for the $N$ cycle dynamics is substituted with the $(N+1)$-th measurement described with $\{M_{y_{N+1},z_{N+1}}\}_{y_{N+1},z_{N+1}}$ (the outcome becomes $b_{N+1} = z_{N+1}$).

\subsection{Implementation of unitary gates and parameter determination} \label{Sss:gate_parameter}
The SiV spin state is manipulated through the application of MW pulses that incur spin rotations via an on-chip gold coplanar waveguide. The MW pulses are near-resonant with the spin MW transition with a detuning of $\Delta_{\omega}$. The phase of the MW pulses sets the axis of the rotation, while the amplitude and duration of the pulses set the angle of the rotation. The MW chain is described in Ref.~\cite{stas2022} of the main text. In the feedback process of $\hat{R}_x(\theta_n)$, an FPGA synchronizes the application of pulses and decides the value $\theta_n$ of the feedback pulse.
Spin rotations $\mathcal{U}_{\varphi}^x$ and $\mathcal{U}_{-\varphi}^x$ included in the heat-bath interaction process $\mathcal{N}$ is implemented in the same way while the amplitude and duration are changed from the $\pi$-rotations.

On the other hand, in the absence of external driving, the SiV spin obeys a Hamiltonian of the form $\Delta_\omega \hat{\sigma}_z$.  Effectively, this will incur a rotation on the spin of the form $\hat{R}_z(\phi)$ with $\phi = \Delta_\omega \Delta t_\mathrm{prec}$, where $\Delta t_\mathrm{prec}$ denotes the precession time. During the precession, the dephasing $\mathcal{D}$ occurs due to the interaction with the environment, where its parameter $\alpha$ is represented as $\alpha \equiv \frac{\Delta t_\mathrm{prec}}{T^*_2}$ with the coherence time $T_2^*$.

\begin{figure*}[]
\begin{center}
\includegraphics[width=0.6\textwidth]{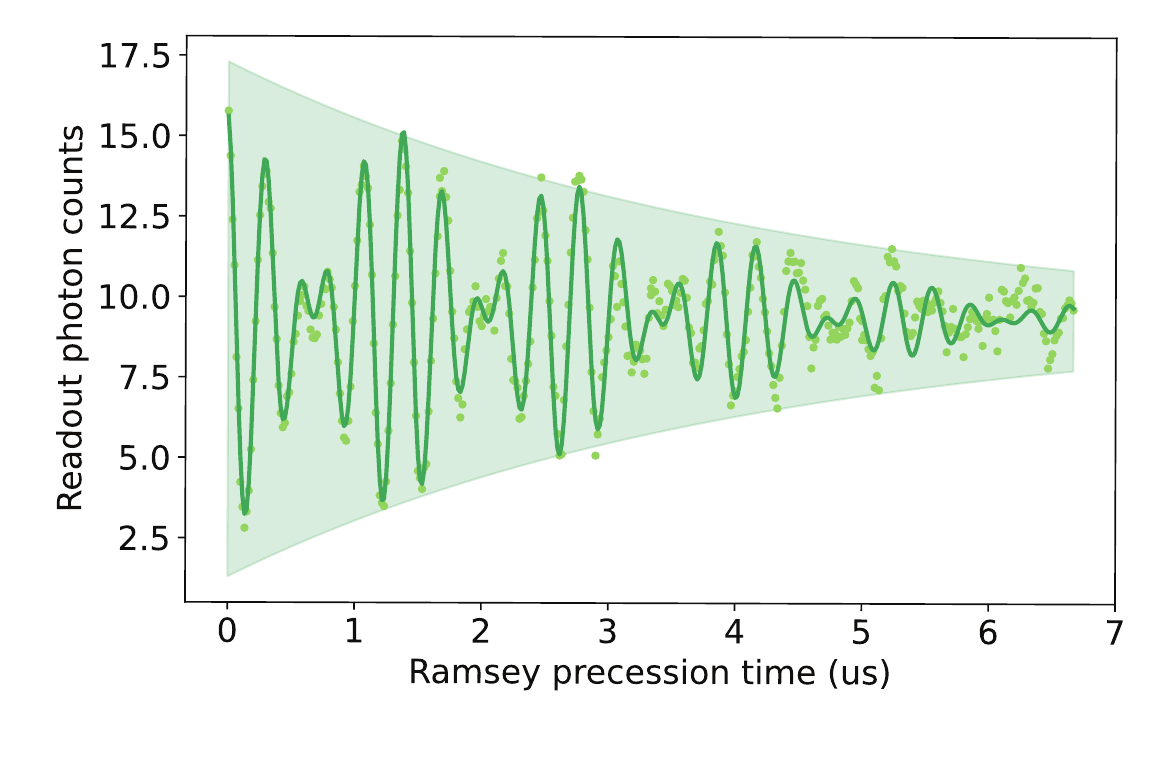}
\caption{Ramsey sequence of the SiV. The dark green solid line is a fit parameterized by $T_2^* = 4.08 \mu\mathrm{s}$ and $\Delta_\omega = \pm 2\pi \times 3.23 \mathrm{MHz}$. The shaded region corresponds to an exponential decay with $T_2^* = 4.08 \mu\mathrm{s}$.}
\label{fig:SI_ramsey}
\end{center}
\end{figure*}

To determine $\Delta_\omega$ and $T_2^*$ and characterize the heat-bath interaction of the spin system, we perform a Ramsey measurement of the spin. This gives us the coherence time $T_2^* = 4.08 \mu \mathrm{s}$. Due to a nearby $^{13}C$ nuclear spin, we see that there are two values of $\Delta_\omega$ dependent on the spin state of the $^{13}C$. By placing the MW drive frequency halfway between the two $^{13}C$ spin state-dependent SiV transition frequencies, we ensure that  and $\Delta_\omega = \pm 2\pi \times 3.23 \mathrm{MHz}$, and thus the absolute value $|\Delta_\omega| = 2\pi \times 3.23 \mathrm{MHz}$ reduces to a single value (Fig. \ref{fig:SI_ramsey}).
Since $\Delta t_\mathrm{prec}$ can be measured in experiment ($\Delta t_\mathrm{prec} = 43.3 \rm ns$ for experiment in Fig.~\ref{fig:MarkovExp} and $\Delta t_\mathrm{prec} = 41.7 \rm ns$ for experiment in Fig.~\ref{fig:nonMarkovExp_GSL}), we can determine the parameters $\phi$ and $\alpha$.

Finally, the parameter $\varphi$ is determined through the fitting of the transition probability $\bra{1} \mathcal{N}(\ket{0}\bra{0}) \ket{1}$. Since the channel $\mathcal{N}\equiv \mathcal{U}_{-\varphi}^x \circ \mathcal{U}_{\phi}^z \circ \mathcal{D} \circ \mathcal{U}_{\varphi}^x$ depends on the rotation angle $\varphi$, we can determine $\varphi$ with $\bra{1} \mathcal{N}(\ket{0}\bra{0}) \ket{1}$ calculated from the experimental data. We can calculate it from sampled data as 
\begin{equation*}
    \bra{1} \mathcal{N}(\ket{0}\bra{0}) \ket{1} = \frac{\mathrm{NUM}(z_n=0,z_{n+1}=1,\theta_n=0)}{\mathrm{NUM}(z_n=0,z_{n+1}=0,\theta_n=0) + \mathrm{NUM}(z_n=0,z_{n+1}=1,\theta_n=0)},
\end{equation*}
where $\mathrm{NUM}(z_n,z_{n+1},\theta_n)$ represents the number of the sampled trajectories with measurement outcomes $(z_n,z_{n+1})$ and feedback angle $\theta_n$. All parameters are summarized in Tab. \ref{table:si_exp_para}.

\begin{table}
\centering
\begin{tabular}{c c c c c c} 
   & Markovian feedback   & \multicolumn{4}{ c }{ Non-Markovian feedback}  \\ 
    &  & $k=1$ &  $k=2$ &  $k=3$ & $k=4$ \\  \hline
   $(\delta_0,\delta_1)$ & $(0.089,0.430)$ & $(0.151,0.505)$ & $(0.157,0.513)$ & $(0.155,0.518)$ & $(0.162,0.516)$ \\
   $\Delta t_\mathrm{prec}$ & $43.3 \rm ns$ & \multicolumn{4}{ c }{$41.7 \rm ns$} \\
   Repetitions & 7000 & \multicolumn{4}{ c }{10000} \\
   $\varphi$ & $0.133\pi$ & \multicolumn{4}{ c }{$0.151\pi$} \\  \hline
   Detuning $|\Delta_{\omega}|$ & \multicolumn{5}{ c }{$2\pi \times 3.23 \mathrm{MHz}$} \\
   Rabi frequency $\Omega_{\omega}$ & \multicolumn{5}{ c }{$2\pi \times 16.7 \mathrm{MHz}$} \\
   $T_2^*$ & \multicolumn{5}{ c }{$4.08 \mu \mathrm{s}$} \\
   $T_1$ & \multicolumn{5}{ c }{$> 1 \mathrm{s}$}
\end{tabular}
\caption{Summary of experimental parameters for Fig. \ref{fig:MarkovExp}, \ref{fig:nonMarkovExp}, and \ref{fig:nonMarkovExp_GSL}.}
\label{table:si_exp_para}
\end{table}

\section{Experiment-numerics hybrid verification method}
In order to verify the generalized FTs (\ref{eq:GFT}) and (\ref{eq:GFT_BQC}) solely with experimental sampling, we need to realize the fine unraveling and $k$-th-order coarse unraveling, respectively, in an experimental platform. However, it is usually not realistic to sample the whole trajectory $\psi_N$. Therefore, in this work, we employ the experiment-numerics hybrid verification method, where we perform the auxiliary numerical calculation in parallel with the experimental trajectory sampling. While this method has already been introduced in Ref.~\cite{yada2022quantum} in the main text, we review it in this section for the completeness of the paper. We overview its essential idea in Sec.~\ref{Sss:hybrid_overview}, and elaborate on the details of numerical analysis in Sec.~\ref{Sss:num_ana}.

\subsection{Overview}\label{Sss:hybrid_overview}
\begin{figure*}[]
\begin{center}
\includegraphics[width=0.75\textwidth]{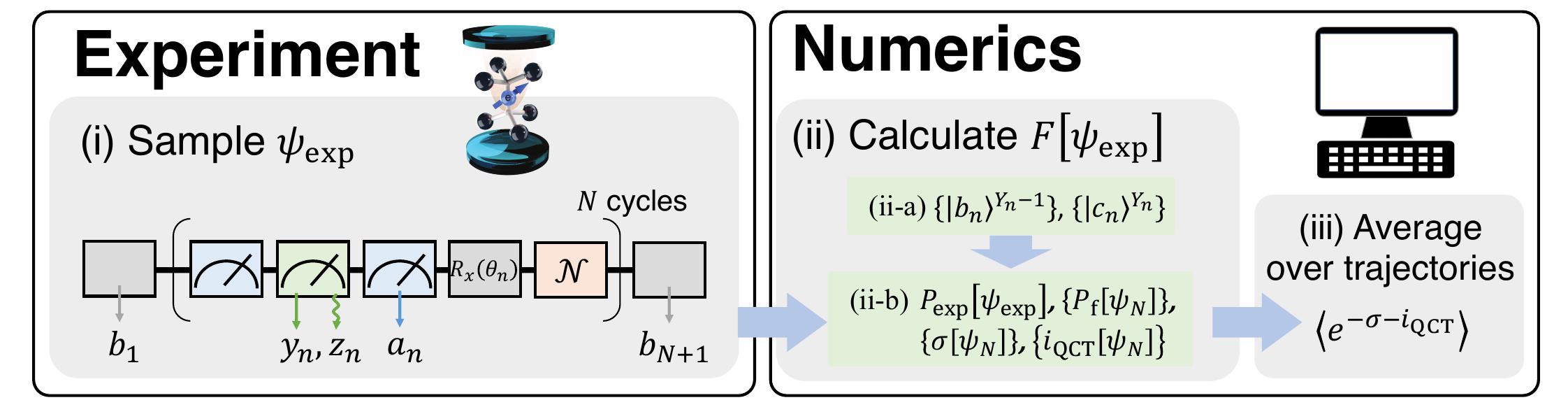}
\caption{Schematic of the experiment-numerics hybrid verification method. To evaluate the averaged value $\langle e^{-\sigma -i_{\rm QCT}} \rangle$, we follow three steps: $(\mathrm{i})$ experimentally sampling the trajectories $\psi_{\rm exp}$, $(\mathrm{ii})$ numerically calculating the quantity $F[\psi_{\rm exp}]$ for each sampled trajectory, and $(\mathrm{iii})$ calculating their average. While this figure focuses on the evaluation of $\langle e^{-\sigma -i_{\rm QCT}} \rangle$, the other quantities plotted in Figs.~\ref{fig:MarkovExp} and \ref{fig:nonMarkovExp_GSL} in the main text are obtained in a similar way.}
\label{fig:SI_expnum_hyb}
\end{center}
\end{figure*}

In the experimental sequence shown in Fig.~\ref{fig:SI_experimental_sequence}, we can obtain the labels $(b_1, \{y_n\}_{n=1}^N, \{z_n\}_{n=1}^N, \{a_n\}_{n=1}^{N},b_{N+1})$. This is because we can optically readout $\{y_n\}_{n=1}^N$ and $\{z_n\}_{n=1}^{N+1}$, and the relationships $b_1 = z_1$, $a_n=z_n$, and $b_{N+1}=z_{N+1}$ are satisfied in this experiment. On the other hand, the other labels $(\{b_n\}_{n=2}^{N}, \{d_n\}_{n=1}^N)$ cannot be obtained experimentally. In such a way, whole trajectory $\psi_N$ can be divided into the experimentally accessible and inaccessible parts as $\psi_N = (\psi_{\rm exp},\psi_{\rm num})$ with $\psi_{\rm exp} \equiv (b_1, \{y_n\}_{n=1}^N, \{z_n\}_{n=1}^N, \{a_n\}_{n=1}^{N},b_{N+1})$ and $\psi_{\rm num} \equiv (\{b_n\}_{n=2}^{N}, \{d_n\}_{n=1}^N)$.

The experiment-numerics hybrid method \cite{yada2022quantum} is used to quantify the ensemble average by combining the experimental sampling of $\psi_{N,\rm exp}$ and auxiliary numerical calculation. For example, we can calculate $\langle e^{-\sigma-i_{\rm QCT}} \rangle$ through the following procedure, as shown in Fig.~\ref{fig:SI_expnum_hyb}:
\begin{itemize}
    \item[$(\mathrm{i})$] Experimentally sample $M$ trajectories $\{\psi_{\rm exp}^i\}_{i=1}^M$.
    \item[$(\mathrm{ii})$] Numerically calculate the quantity $F[\psi_{\rm exp}^i] \equiv \sum_{\psi_{\rm num}} \frac{P_{\rm f}[\psi_{\rm exp}^i,\psi_{\rm num}]}{P_{\rm exp}[\psi_{\rm exp}^i]}e^{-\sigma[\psi_{\rm exp}^i,\psi_{\rm num}]-i_{\rm QCT}[\psi_{\rm exp}^i,\psi_{\rm num}]}$ for each sampled trajectory $\psi_{\rm exp}^i$.
    \item[$(\mathrm{iii})$] Average $F[\psi_{\rm exp}^i]$ over all the sampled trajectories.
\end{itemize}
The details of the numerical calculation part $(\mathrm{ii})$ are discussed in Sec.~\ref{Sss:num_ana}.
Since the probability to sample $\psi_{\rm exp}$ follows $P_{\rm exp}[\psi_{\rm exp}]$, the averaged value taken in $(\mathrm{iii})$, denoted as $\mathbb{E}[F] \equiv \sum_{i=1}^M \frac{F[\psi_{\rm exp}^i]}{M}$, would converge to $\langle e^{-\sigma-i_{\rm QCT}} \rangle$ by increasing the number of sampled trajectories $M$.
\begin{equation*}
    \mathbb{E}[F]  \longrightarrow  \langle e^{-\sigma-i_{\rm QCT}} \rangle \equiv \sum_{\psi_N} P_{\rm f}[\psi_N]e^{-\sigma[\psi_N]-i_{\rm QCT}[\psi_N]}
\end{equation*}
Thus, we can evaluate $\langle e^{-\sigma-i_{\rm QCT}} \rangle$ from the experimentally sampled data.
The other averaged quantities shown in Figs.~\ref{fig:MarkovExp} and \ref{fig:nonMarkovExp_GSL} in the main text are also evaluated in a similar way as $\langle e^{-\sigma - i_{\rm QCT}}\rangle$.

\subsection{Numerical analysis}\label{Sss:num_ana}
The numerical calculation part $(\mathrm{ii})$ of the hybrid method is composed of two steps: $(\mathrm{ii-a})$ calculating the basis sets $\{\ket{b_n}^{Y_{n-1}}\}$ and $\{\ket{a_n}^{Y_{n}}\}$ used in the fine unraveling, and $(\mathrm{ii-b})$ evaluating $F[\psi_{\rm exp}]$ through trajectory-level calculations.

In $(\mathrm{ii-a})$, we calculate the conditional density operators $\rho_n^{Y_{n-1}}$ and $\tau_n^{Y_n}$ to obtain
$\{\ket{b_n}^{Y_{n-1}}\}$ and $\{\ket{a_n}^{Y_n}\}$, respectively. 
In the current sequence, we can immediately show that $\tau_n^{Y_n}$ is diagonal with respect to $\{\ket{0},\ket{1}\}$ for all $n$ and $Y_n$, from the definition of the measurement operators $\{M_{y_n,z_n}\}_{y_n,z_n}$. Therefore $\{\ket{a_n}^{Y_n}\}$ becomes $\{\ket{0},\ket{1}\}$ independent of $n$ and $Y_n$. 
Furthermore, we can show that the diagonal basis of $\rho_n^{Y_{n-1}}$ is also fixed and independent of $n$ or $Y_{n-1}$ from the following equation:
\begin{equation}
    \begin{split}
        \rho^{Y_{n-1}}_n &= \mathcal{N}(\hat{R}_x(\theta_{n-1}) \tau_{n-1}^{Y_{n-1}} \hat{R}_x(\theta_{n-1})^\dag ) \\
    &= \mathcal{N}(p(a_{n-1}=0|Y_{n-1}) \ket{0}\bra{0} + p(a_{n-1}=1|Y_{n-1})\ket{1}\bra{1}) \\
    &= p(0|Y_{n-1})\mathcal{N}(\mathbbm{1}) + (1-2p(0|Y_{n-1}))\mathcal{N}(\ket{1}\bra{1}) \\
    &= p(0|Y_{n-1})\mathbbm{1} + (1-2p(0|Y_{n-1}))\mathcal{N}(\ket{1}\bra{1}),
    \end{split} \label{eq:SI_fine_coarse_equiv}
\end{equation}
where we use in the final equality that the maximally mixed state $\mathbbm{1}$ is the fixed point of the quantum channel $\mathcal{N}$. As is clear from this equation, the basis $\{\ket{b_n}^{Y_{n-1}}\}$ is the diagonal basis of $\mathcal{N}(\ket{1}\bra{1})$, which is described as $\{\ket{\xi_0},\ket{\xi_1}\}$.
It would be noteworthy that in this experiment, the fine unraveling and $k$-th-order coarse unraveling become equivalent since the inserted diagonal bases $\{\ket{b_{n+1}}^{Y_{n}}\}$ and $\{\ket{a_n}^{Y_{n}}\}$ are independent of $Y_n$.

In $(\mathrm{ii-b})$, we calculate $F[\psi_{\rm exp}^i]$ for each sampled trajectory $\psi_{\rm exp}^i$ by utilizing the diagonal bases calculated in $(\mathrm{ii-a})$. 
For that purpose, we need to quantify $\{P_{\rm f}[\psi_{\rm exp}^i,\psi_{\rm num}]\}_{\psi_{\rm num}}$, $\{\sigma[\psi_{\rm exp}^i,\psi_{\rm num}]\}_{\psi_{\rm num}}$, and $\{i_{\rm QCT}[\psi_{\rm exp}^i,\psi_{\rm num}]\}_{\psi_{\rm num}}$ for all possible $\psi_{\rm num}$, and $P_{\rm exp}[\psi_{\rm exp}]$.
The quantities $P_{\rm f}[\psi_N]$, $\sigma[\psi_N]$, and $i_{\rm QCT}[\psi_N]$ are defined in Appendix~\ref{appss:thermo_QCTE_GFT}, and $P_{\rm exp}[\psi_{\rm exp}]$ is defined as follows:
\begin{equation*}
    P_{\rm exp}[\psi_{\rm exp}] = \bra{b_{N+1}} \mathcal{L}_{N}^{Y_N} (\Pi_{a_N}^{Y_N}) \ket{b_{N+1}} \left(\prod_{n=2}^{N} \left\| \Pi_{a_{n}}^{Y_{n}} M_{y_n,z_n} \mathcal{L}_{n-1}^{Y_{n-1}} (\Pi_{a_{n-1}}^{Y_{n-1}}) M_{y_n,z_n}^\dag \Pi_{a_{n}}^{Y_{n}} \right\|^2 \right) \left\|\Pi_{a_1}^{y_1}M_{y_1,z_1}\ket{b_1} \right\|^2 p(b_1),
\end{equation*}
where $\Pi_{a_{n}}^{Y_{n}}\equiv \ket{a_n}\bra{a_n}^{Y_n}$ is the rank-one projector.

\section{Additional experimental data}\label{Ss:extra_nonmarkov}
\subsection{Markovian feedback experiment}

\begin{figure*}[]
\begin{center}
\includegraphics[width=0.65\textwidth]{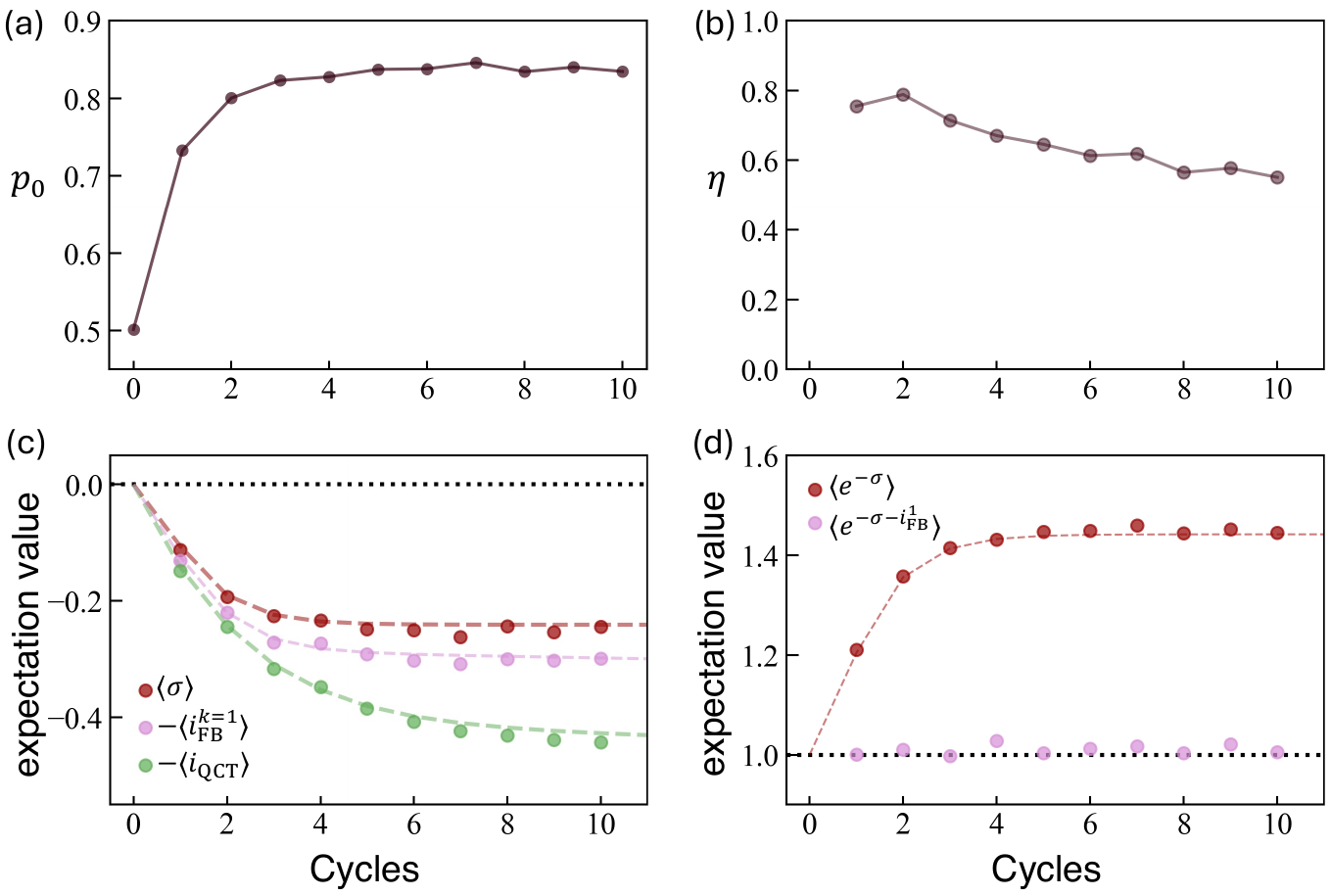}
\caption{Additional experimental data taken with Markovian feedback.
(a,b) Experimentally measured ground-state population $p_0$ and efficiency $\eta$ as a function of the number of feedback cycles.
(c) Experimentally obtained entropy production $\langle \sigma \rangle$, negative QC-transfer entropy $-\langle i_{\rm QCT}\rangle$, and negative value of the net information fed back through Markovian feedback $-\langle i_{\rm FB}^{k=1}\rangle$ against the number of feedback cycles. (d) Statistical averages $\langle e^{-\sigma}\rangle$ and $\langle e^{-\sigma -i_{\rm FB}^{k=1}}\rangle$ from experimental data are plotted against the number of feedback cycles. 
Dotted lines in Figs. (c) and (d) represent theoretical predictions for the same colored plots, obtained from numerical simulations using the experimental parameters. In all figures, the error bars are omitted, since the standard error of the mean are smaller than the marker sizes.
}
\label{fig:SI_markov_i_FB}
\end{center}
\end{figure*}

Figure~\ref{fig:SI_markov_i_FB} presents additional experimental data for the Markovian feedback experiment discussed in Sec.~\ref{ss:Markov_exp} of the main text.  
Figures~\ref{fig:SI_markov_i_FB}(a) and (b) show the ground-state population and the conversion efficiency as functions of the number of feedback cycles.  
Figures~\ref{fig:SI_markov_i_FB}(c) and (d) provide data that can be used to verify the generalized SL and FT. Specifically, they show the dynamics of $-\langle i_{\rm FB}^{k=1}\rangle$ and $\langle e^{-\sigma -i_{\rm FB}^{k=1}}\rangle$, respectively, which were not included in Fig.~\ref{fig:MarkovExp} in the main text.  
These results confirm that the generalized SL, $\langle \sigma \rangle \geq -\langle i_{\rm FB}^{k=1}\rangle\geq -\langle i_{\rm QCT}\rangle$, and FT, $\langle e^{-\sigma -i_{\rm FB}^{k=1}}\rangle = 1$, are both satisfied.

\subsection{Non-Markovian feedback experiment}

\begin{figure*}[]
\begin{center}
\includegraphics[width=0.9\textwidth]{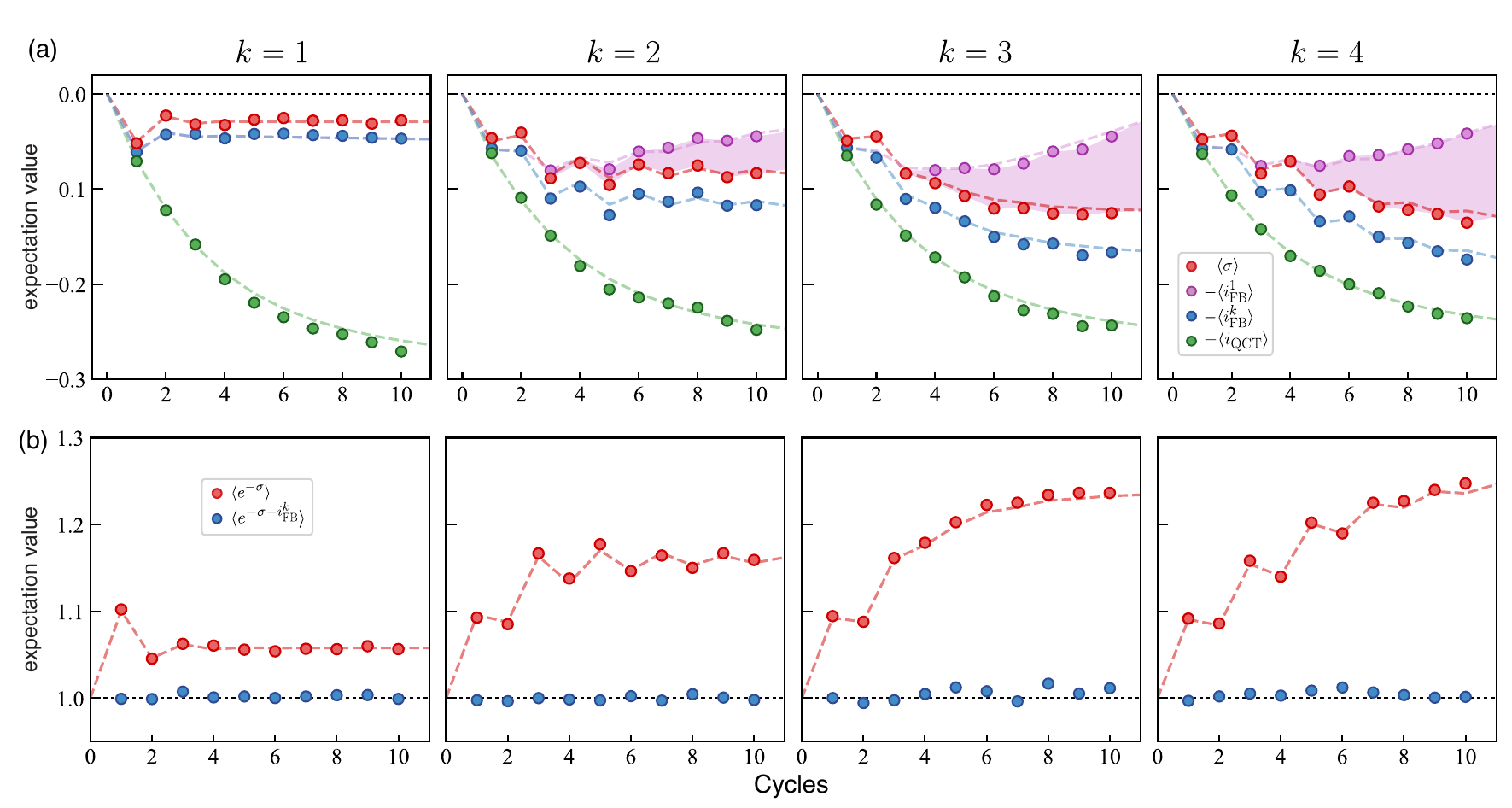}
\caption{Additional experimental data for the non-Markovian feedback experiment shown with $k$ = 1, 2, 3, and 4.(a) Experimentally measured entropy production $\langle \sigma \rangle$, lower bound for the entropy production under Markovian feedback $-\langle i_{\rm FB}^{1} \rangle$, negative values of the net information fed back through $k$-th-order Markovian feedback $-\langle i_{\rm FB}^{k} \rangle$, and negative QC-transfer entropy $-\langle i_{\rm QCT} \rangle$ are plotted against the number of feedback cycles. (b) Experimentally measured $\langle e^{-\sigma}\rangle$ and $\langle e^{-\sigma -i_{\rm FB}^k}\rangle$ are plotted against the number of feedback cycles. In both (a) and (b), dotted lines represent theoretical predictions corresponding to the same colored plots.
Error bars are omitted in all figures since the standard error of the mean are smaller than the marker sizes.}
\label{fig:SI_extra_nonmarkovian_data}
\end{center}
\end{figure*}

Figure \ref{fig:SI_extra_nonmarkovian_data} shows additional experimental data taken with non-Markovian feedback with different values of $k$ ranging from 1 to 4 (figures of $k=1$ and $4$ are included in Fig.~\ref{fig:nonMarkovExp_GSL} in the main text). We can confirm that the generalized SL (\ref{eq:GSL_BQC}) and FT (\ref{eq:GFT_BQC}) are satisfied for all $k$. Larger entropy reduction is realized with higher-order Markovian feedback. The shaded purple region, which represents the thermodynamic advantage of non-Markovian feedback over Markovian feedback, is also larger for higher $k$.

\begin{figure*}[]
\begin{center}
\includegraphics[width=0.9\textwidth]{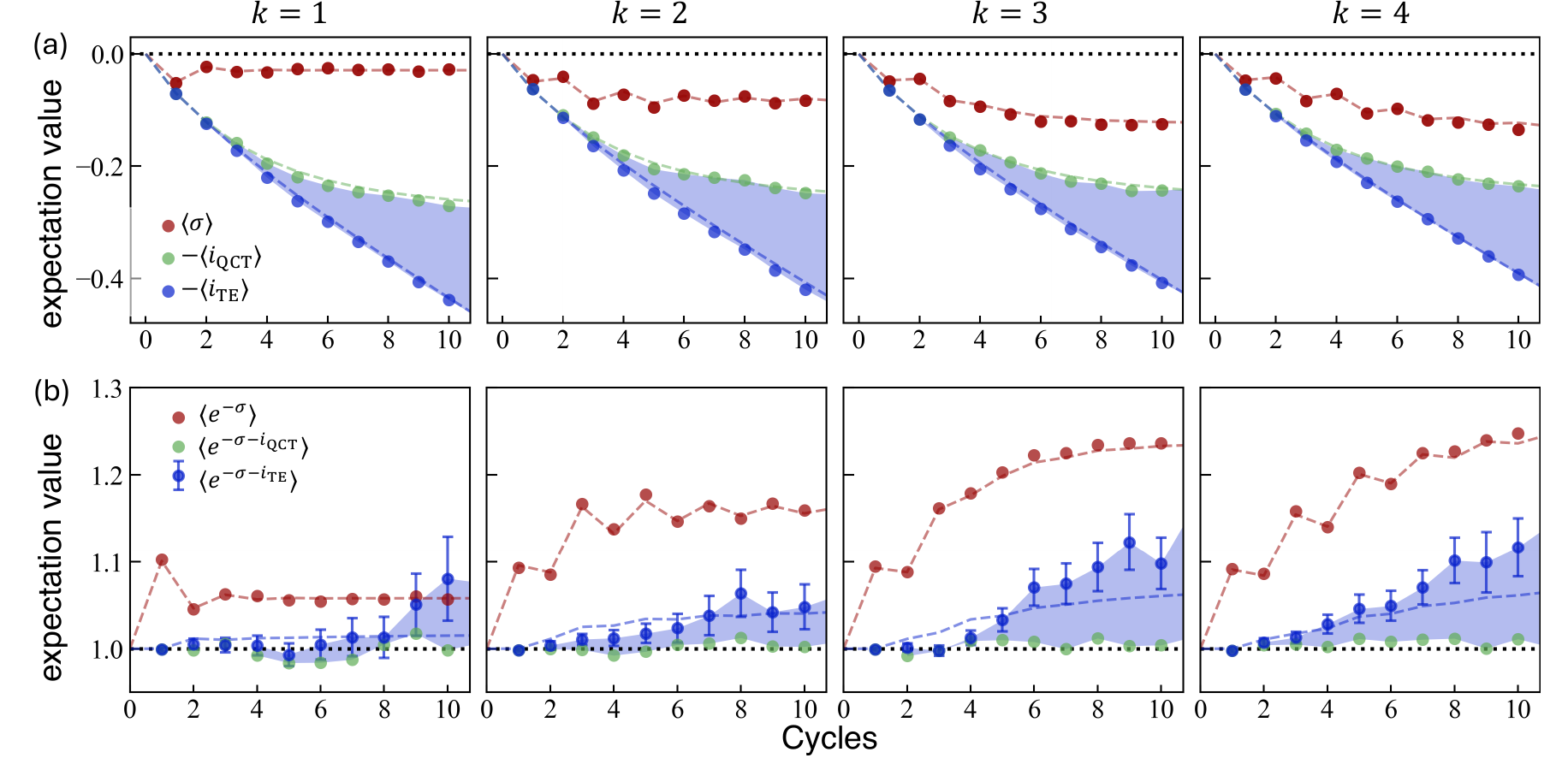}
\caption{
Additional experimental data for the non-Markovian feedback experiment including the (classical) transfer entropy $i_{\rm TE}$. 
(a) Experimentally measured entropy production $\langle \sigma \rangle$, negative QC-transfer entropy $-\langle i_{\rm QCT} \rangle$, and negative transfer entropy $-\langle i_{\rm TE} \rangle$ are plotted against the number of feedback cycles. 
(b) Experimentally measured $\langle e^{-\sigma}\rangle$, $\langle e^{-\sigma -i_{\rm QCT}}\rangle$, and $\langle e^{-\sigma -i_{\rm TE}}\rangle$ are plotted against the number of feedback cycles. In both (a) and (b), dotted lines represent theoretical predictions corresponding to the same colored plots. Error bars, which represent the standard error of the mean, are omitted except for $\langle e^{-\sigma -i_{\rm TE}}\rangle$, since they are smaller than the marker sizes.
}
\label{fig:SI_nonmarkov_CTE}
\end{center}
\end{figure*}

Figure~\ref{fig:SI_nonmarkov_CTE} shows additional experimental results for the non-Markovian feedback experiment, including the behavior of the classical transfer entropy $i_{\rm TE}$. From Fig.~\ref{fig:SI_nonmarkov_CTE}(a), we observe that the classical transfer entropy and the QC-transfer entropy take different values due to the effect of measurement backaction. Figure~\ref{fig:SI_nonmarkov_CTE}(b) demonstrates that while the generalized FT incorporating the QC-transfer entropy holds, i.e., $\langle e^{-\sigma - i_{\rm QCT}} \rangle \sim 1$, the corresponding expression using the classical transfer entropy, $\langle e^{-\sigma - i_{\rm TE}} \rangle$, deviates significantly from 1.
This deviation is particularly prominent for $k=3$ and $k=4$, whereas for $k=1$ and $k=2$, the values remain mostly within the error bars.
This larger deviation at $k=3$ and $k=4$ arises because the entropy reduction becomes greater at higher values of $k$.

\section{Additional numerical data}

\begin{figure*}[]
\begin{center}
\includegraphics[width=0.6\textwidth]{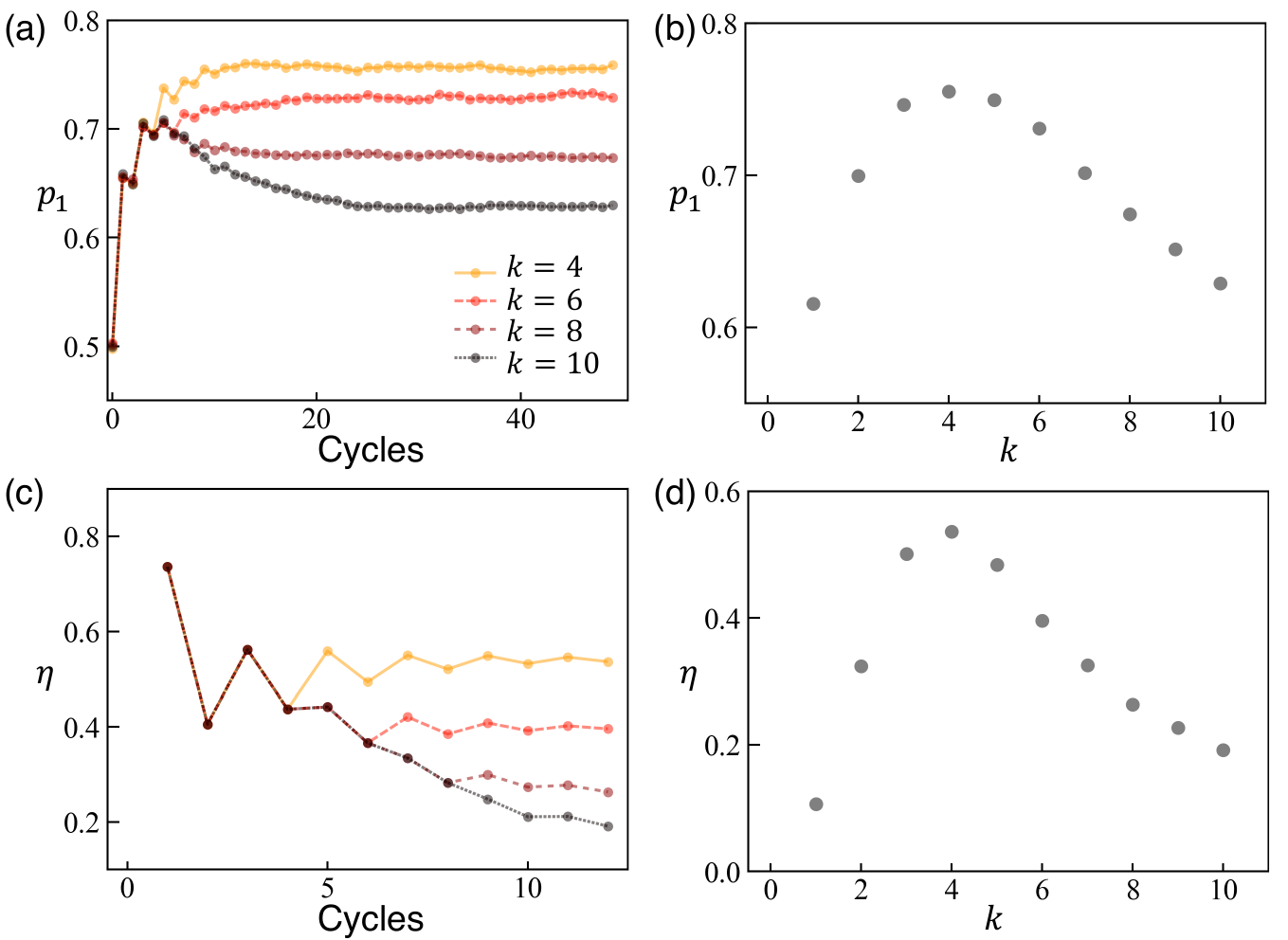}
\caption{Numerical results for $k$-th-order Markovian feedback with memory lengths $1 \leq k \leq 10$, using the same parameters as those in the experiment. (a), (b) Time evolution and steady-state values of the excited-state population $p_1$. For (a), the values are obtained by numerically sampling $50{,}000$ trajectories and taking their average. The steady-state values in (b) are calculated by averaging $p_1$ over the final 10 cycles ($n \in [41, 50]$). (c), (d) Time evolution and values at $n = 12$ of the conversion efficiency $\eta$. These quantities are calculated exactly by computing the conditional density operators $\tau_n^{Y_n}$ and $\rho_{n+1}^{Y_n}$ for all $n,Y_n$. We restrict the calculation to $n \leq 12$ due to the exponential growth of computational cost with $n$. The parameters used in all panels (a–d) are summarized in the “Non-Markovian feedback” column of Table~\ref{table:si_exp_para}. For consistency, $(\delta_0, \delta_1)$ is taken as the average value $(0.156, 0.513)$.
}
\label{fig:SI_nonmarkov_k_dependence}
\end{center}
\end{figure*}

Figure~\ref{fig:SI_nonmarkov_k_dependence} shows the results of numerical calculations using the same parameters as those employed in the non-Markovian feedback experiment. In Figs.~\ref{fig:SI_nonmarkov_k_dependence}(a) and (b), the dynamics and the steady-state values of the excited-state population $p_{1}$ are shown for different $k$. These figures indicate that the excited state is most stable at $k = 4$, and that higher-order Markovian feedback beyond this point worsens the stability. A similar trend is observed for the conversion efficiency $\eta$ in Figs.~\ref{fig:SI_nonmarkov_k_dependence}(c) and (d).

The reason for this performance degradation at larger $k$ lies in the specific feedback protocol employed in this work. In the current protocol, a $\pi$-pulse is applied only when the last $k$ outcomes are all zero, which is just one of the $2^k$ possible outcome patterns. Therefore, as $k$ increases, $(\mathrm{i})$ the failure probability to convert $\ket{1}$ to $\ket{0}$ decreases due to the larger memory size, while $(\mathrm{ii})$ the chance to convert $\ket{0}$ to $\ket{1}$ also decreases. Because of these two competing effects, the optimal value of $k$ becomes $k = 4$ in this case.

However, it should be noted that the second effect discussed here is specific to the set of feedback protocols employed in this work and does not represent a general trend for $k$-th-order Markovian feedback. If the optimal feedback protocol were chosen from the entire set of $k$-th-order Markovian protocols, the performance would improve with increasing $k$, since the set of $k$-th-order Markovian feedback protocols is included in that of $(k+1)$-th-order protocols by definition.

\section{Derivation of Eq.~(\ref{eq:GFT_BQC})} \label{Sss:GFT_BQC}
\begin{figure*}[]
\begin{center}
\includegraphics[width=0.80\textwidth]{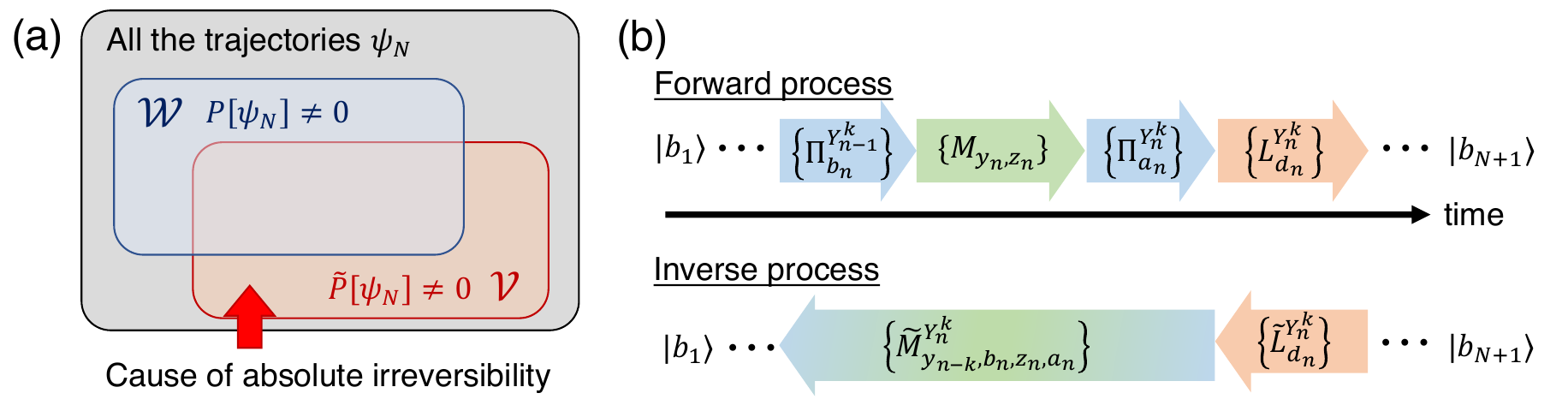}
\caption{Derivation of the generalized fluctuation theorem under $k$-th-order coarse unraveling. (a) Venn diagram for the set of trajectories under $k$-th-order coarse unraveling. $\mathcal{W}$ denotes the set of labels $\psi_N$ for which $P_{k,\rm c}[\psi_N] \neq 0$, while $\mathcal{V}$ is the set of $\psi_N$ satisfying $\Tilde{P}_{k,\rm c}[\psi_N] \neq 0$. The trajectories in $\mathcal{V} \setminus \mathcal{W}$ are the cause of absolute irreversibility. (b) Schematics of the forward process and inverse process. The artificial inverse process consists of two parts: the inverse of the measurement process and inserted PMs, which is represented with Kraus operators $\{\Tilde{M}_{y_{n-k},b_{n},z_n,a_{n}}^{Y_n^k}\}_{y_{n-k},b_{n},z_n,a_{n}}$, and the inverse of the feedback process, where the Kraus operators are $\{\Tilde{L}_{d_{n}}^{Y_n^k}\}_{d_n}$.}
\label{fig:SI_GFT_derivation}
\end{center}
\end{figure*}

We here provide the derivation of the generalized FT under the $k$-th-order coarse unraveling,
\begin{equation*}
    \langle e^{-\sigma-i_{\rm FB}^k} \rangle = 1.
\end{equation*}
The essence of the derivation is basically the same as that of Eq.~(\ref{eq:GFT}) discussed in Ref.~\cite{yada2022quantum}: we introduce an artificial stochastic process called inverse process, in which each trajectory is labeled by $\psi_N$.
The inverse process is required to satisfy the following two conditions, namely the generalized detailed FT~(\ref{eq:GDFT}) and probability normalization condition~(\ref{eq:prob_norm}):
\begin{gather}
    \frac{\Tilde{P}_{k,\rm c}[\psi_N]}{P_{k,\rm c}[\psi_N]} = e^{-\sigma[\psi_N]-i_{\rm FB}^k[\psi_N]}, \label{eq:GDFT} \\
    \sum_{\psi_N} \Tilde{P}_{k,\rm c}[\psi_N] = 1, \label{eq:prob_norm}
\end{gather}
where $\Tilde{P}_{k,\rm c}[\psi_N]$ denotes the probability for the trajectory $\psi_N$ in the inverse process. 
By utilizing this inverse process, we can derive the generalized FT as
\begin{equation}
\begin{split}
    \langle e^{-\sigma-i_{\rm FB}^k} \rangle &= \sum_{\psi_N \in \mathcal{W}} P_{k,\rm c}[\psi_N]\frac{\Tilde{P}_{k,\rm c}[\psi_N]}{P_{k,\rm c}[\psi_N]} \\
    &= \sum_{\psi_N \in \mathcal{W}}\Tilde{P}_{k,\rm c}[\psi_N] \\
    &= 1- \lambda_{\rm irr},
\end{split}
\end{equation}
where the absolute irreversibility term $\lambda_{\rm irr}$ is defined as
\begin{equation}
\label{eq:abs_irr}
    \lambda_{\rm irr} \equiv \sum_{\psi_N \in  (\mathcal{V} \smallsetminus \mathcal{W})} \Tilde{P}_{k,\rm c}[\psi_N].
\end{equation}
Here, $\mathcal{W}$ and $\mathcal{V}$ are defined as the set of labels $\psi_N$, which satisfies $P_{k,\rm c}[\psi_N] \neq 0$ and $\Tilde{P}_{k,\rm c}[\psi_N] \neq 0$, respectively, as shown in Fig.~\ref{fig:SI_GFT_derivation}(a).
In the rest of this section, we first introduce the inverse process which satisfies Eqs.~(\ref{eq:GDFT}) and (\ref{eq:prob_norm}), and next derive the sufficient condition for the absolute irreversibility term to be zero (i.e., $\lambda_{\rm irr} = 0$).

The inverse process for the $k$-th-order coarse unraveling is realized by applying two types of CPTP map alternately for $N$ times, as shown in Fig.~\ref{fig:SI_GFT_derivation}(b).
The first one corresponds to the feedback process $\mathcal{L}^{Y_n^k}_n$ of the original (forward) dynamics. It is described with the set of Kraus operators $\{\Tilde{L}_{d_{n}}^{Y_n^k}\}_{d_n}$, which is defined as
\begin{equation}
\label{Kraus_ref_diss}
    \Tilde{L}_{d_{n}}^{Y_n^k} \equiv L_{d_{n}}^{Y_n^k \dag} e^{-\frac{\beta \Delta_{d_n}}{2}}.
\end{equation}
We can prove this set of operators satisfies the trace preserving condition as follows:
\begin{equation}
\label{seq:TP_Rdiss}
\begin{split}
    \sum_{d_n} \Tilde{L}_{d_{n}}^{Y_n^k \dag} \Tilde{L}_{d_{n}}^{Y_n^k} 
    &= \sum_{m,l} L_{(m,l)}^{Y_n^k} L_{(m,l)}^{Y_n^k \dag} e^{-\beta \Delta_{(m,l)}} \\
    &= \sum_{m,l} \bra{E_l}U^{Y_n^k}\ket{E_m} \bra{E_m}U^{Y_n^k\dag}\ket{E_l} \frac{e^{-\beta E_l}}{Z_{\beta}} \\
    &= \mathbbm{1}.
\end{split}
\end{equation}

The second CPTP map corresponds to the $n$-th measurement $\{M_{y_n,z_n}\}_{y_n,z_n}$ and inserted PMs $\{\Pi_{b_n}^{Y_{n-1}^k}\}_{b_n}$ and $\{\Pi_{a_n}^{Y_n^k}\}_{a_n}$. 
For $k+1\leq n \leq N$. It is represented with the set of Kraus operators $\{\Tilde{M}_{y_{n-k},b_{n},z_n,a_{n}}^{Y_n^k}\}_{y_{n-k},b_{n},z_n,a_{n}}$ defined as 
\begin{equation}
\label{Kraus_inverse_fine_detail_M_B}
\Tilde{M}_{y_{n-k},b_{n},z_n,a_{n}}^{Y_n^k}\equiv
\begin{cases}
      \ket{b_{n}}^{Y_{n-1}^k}\  {}^{Y_{n}^k}\bra{a_{n}}\sqrt{\frac{\left|{}^{Y_{n}^k}\bra{a_{n}}M_{y_{n},z_n}\ket{b_{n}}^{Y_{n-1}^k}\right|^2 p(b_{n}|Y_{n-1}^k)P[Y_{n-1}^{k}]}{p(a_{n}|Y_{n}^{k})P[Y_{n}^{k}]}}
      &(\mathrm{if}\quad p(a_{n}|Y_{n}^{k})P[Y_{n}^{k}] \neq 0)\\
      \ket{b_{n}}^{Y_{n-1}^k}\  {}^{Y_{n}^k}\bra{a_{n}} \frac{1}{\sqrt{D_{n}}} 
      &(\mathrm{if}\quad p(a_{n}|Y_{n}^{k})P[Y_{n}^{k}]  = 0)\\
\end{cases},
\end{equation}
where $D_n \equiv \sum_{y_{n-k},b_{n},z_n} 1$ represents the total number of the label set $(y_{n-k},b_{n},z_n)$.
For $1 \leq n\leq k$, it is described with the Kraus operators $\{\Tilde{M}_{b_{n},z_n,a_{n}}^{Y_n}\}_{b_{n},z_n,a_{n}}$, defined as 
\begin{equation}
\label{Kraus_inverse_fine_detail_M_B_final}
\Tilde{M}_{b_{n},z_n,a_{n}}^{Y_n}\equiv
\begin{cases}
      \ket{b_{n}}^{Y_{n-1}}\  {}^{Y_{n}}\bra{a_{n}} \sqrt{\frac{\left|{}^{Y_{n}}\bra{a_{n}}M_{y_{n},z_n}\ket{b_{n}}^{Y_{n-1}} \right|^2 p(b_{n}|Y_{n-1})P[Y_{n-1}]}{p(a_{n}|Y_{n})P[Y_{n}]} } &(\mathrm{if}\quad p(a_{n}|Y_{n})P[Y_{n}] \neq 0)\\
      \ket{b_{n}}^{Y_{n-1}}\  {}^{Y_{n}}\bra{a_{n}} \frac{1}{\sqrt{D_{n}}}  &(\mathrm{if}\quad p(a_{n}|Y_{n})P[Y_{n}]  = 0)\\
\end{cases},
\end{equation}
with the number of labels $D_n \equiv \sum_{b_{n},z_n} 1$. 
By defining the set of $a_n$ satisfying $p(a_{n}|Y_{n}^{k})P[Y_{n}^{k}] \neq 0$ for fixed $Y_{n}^{k}$ as $\mathcal{A}_{>0}$ and set of $a_n$ satisfying $p(a_{n}|Y_{n}^{k})P[Y_{n}^{k}] = 0$ as $\mathcal{A}_{=0}$, we can derive the trace preserving condition for $\{\Tilde{M}_{y_{n-k},b_{n},z_n,a_{n}}^{Y_n^k}\}_{y_{n-k},b_{n},z_n,a_{n}}$ as follows:
\begin{equation}
\label{seq:TP_Minv}
\begin{split}
    \sum_{\substack{y_{n-k},b_n, \\ z_n,a_n}} \Tilde{M}_{y_{n-k},b_{n},z_n,a_{n}}^{Y_n^k \dag} &\Tilde{M}_{y_{n-k},b_{n},z_n,a_{n}}^{Y_n^k} 
    = \sum_{ \substack{a_n\in \mathcal{A}_{>0} \\ y_{n-k},b_n,z_n} } \Pi_{a_n}^{Y_n^k} \frac{\left|{}^{Y_{n}^k}\bra{a_{n}}M_{y_{n},z_n}\ket{b_{n}}^{Y_{n-1}^k}\right|^2 p(b_{n}|Y_{n-1}^k)P[Y_{n-1}^{k}]}{p(a_{n}|Y_{n}^{k})P[Y_{n}^{k}]} 
    + \sum_{ \substack{a_n\in \mathcal{A}_{=0} \\ y_{n-k},b_n,z_n} } \Pi_{a_n}^{Y_n^k}  \frac{1}{D_n} \\
    &= \sum_{a_n\in \mathcal{A}_{>0}} \Pi_{a_n}^{Y_n^k} 
     \frac{{}^{Y_{n}^k}\bra{a_{n}} \left\{\sum_{y_{n-k}}  P[Y_{n-1}^{k}] \left(\sum_{z_n} M_{y_{n},z_n} \rho_n^{Y_{n-1}^{k}} M_{y_{n},z_n}^\dag\right) \right\} \ket{a_{n}}^{Y_{n}^k} }{p(a_{n}|Y_{n}^{k})P[Y_{n}^{k}]} 
    + \sum_{a_n \in \mathcal{A}_{=0}} \Pi_{a_n}^{Y_n^k} \\
    &= \sum_{a_n\in \mathcal{A}_{>0}} \Pi_{a_n}^{Y_n^k}  + \sum_{a_n\in \mathcal{A}_{=0}} \Pi_{a_n}^{Y_n^k} = \mathbbm{1},
\end{split}
\end{equation}
where the third equality follows from the equation $\sum_{y_{n-k}}  P[Y_{n-1}^{k}] \left(\sum_{z_n} M_{y_{n},z_n} \rho_n^{Y_{n-1}^{k}} M_{y_{n},z_n}^\dag\right) = P[Y_{n}^{k}] \tau_n^{Y_{n}^{k}}$. In a similar way, we can prove the trace-preserving condition for $\{\Tilde{M}_{b_{n},z_n,a_{n}}^{Y_n}\}_{b_{n},z_n,a_{n}}$ for $n\leq k$:
\begin{equation}
\label{seq:TP_Minv_f}
    \sum_{b_{n},z_n,a_{n}} \Tilde{M}_{b_{n},z_n,a_{n}}^{Y_n\dag} \Tilde{M}_{b_{n},z_n,a_{n}}^{Y_n}= \mathbbm{1}.
\end{equation}

With the operators defined in Eqs.~(\ref{Kraus_ref_diss}), (\ref{Kraus_inverse_fine_detail_M_B}), and (\ref{Kraus_inverse_fine_detail_M_B_final}), the total inverse process is described with the following set of Kraus operators:
\begin{equation*}
    \left\{ \left(\prod_{n=k}^{1} \Tilde{M}_{b_{n},z_n,a_{n}}^{Y_n} \Tilde{L}_{d_{n}}^{Y_n} \right) \left( \prod_{n=N}^{k+1} \Tilde{M}_{y_{n-k},b_{n},z_n,a_{n}}^{Y_n^k} \Tilde{L}_{d_{n}}^{Y_n^k} \right) \right\}_{\psi_N \setminus (Y_N^k, b_{N+1}) }
\end{equation*}
By taking the initial state as $\ket{b_{N+1}}$ with probability $P[Y_N^k] p(b_{N+1}|Y_N^k)$, the inverse probability for the trajectory $\psi_N \in \mathcal{W}$ is defined as 
\begin{equation}
    \begin{split}
        \Tilde{P}_{k,\rm c}&[\psi_N] 
        \equiv
\left\|\left(\prod_{n=k}^{1} \Tilde{M}_{b_{n},z_n,a_{n}}^{Y_n} \Tilde{L}_{d_{n}}^{Y_n} \right) \left( \prod_{n=N}^{k+1} \Tilde{M}_{y_{n-k},b_{n},z_n,a_{n}}^{Y_n^k} \Tilde{L}_{d_{n}}^{Y_n^k} \right) \ket{b_{N+1}}\right\|^2 
P[Y_N^k] p(b_{N+1}|Y_N^k)\\
&= \left\{\prod_{n=1}^{N} \left|{}^{Y_n^k}\bra{a_{n}}M_{y_{n},z_n}\ket{b_{n}}^{Y_{n-1}^k}\right|^2
\frac{p(b_{n}|Y_{n-1}^{k})}{p(a_{n}|Y_{n}^{k})}\frac{P[Y_{n-1}^k]}{P[Y_{n}^k]}\left|{}^{Y_n^k}\bra{a_n} \Tilde{L}_{d_{n}}^{Y_n^k} \ket{b_{n+1}}^{Y_{n}^k}\right|^2 \right\}  P[Y_N^k] p(b_{N+1}|Y_N^k) ,
    \end{split}\label{eq:prob_ref}
\end{equation}
where $\ket{b_{N+1}}^{Y_N^k} \equiv \ket{b_{N+1}}$. We here utilize the fact that $p(a_{n}|Y_{n}^{k})P[Y_{n}^{k}] \neq 0$ is satisfied for all $n$ if $\psi_N \in \mathcal{W}$ (i.e., $P_{k,\rm c}[\psi_N] \neq 0$).
For such an inverse process, the probability normalization condition (\ref{eq:prob_norm}) is satisfied due to the trace-preserving conditions, Eqs.~(\ref{seq:TP_Rdiss}), (\ref{seq:TP_Minv}), and (\ref{seq:TP_Minv_f}), and the equality $\sum_{Y_N^k,b_{N+1}}P[Y_N^k] p(b_{N+1}|Y_N^k) = 1$.

We can further prove the other condition, the generalized detailed fluctuation theorem~(\ref{eq:GDFT}), is satisfied for this inverse process, as follows:
\begin{align*}
\frac{\Tilde{P}_{k,\rm c}[\psi_N]}{P_{k,\rm c}[\psi_N]} &= \left \{\prod_{n=1}^{N} \frac{\left|{}^{Y_n^k}\bra{a_n} \Tilde{L}_{d_{n}}^{Y_n^k} \ket{b_{n+1}}^{Y_n^k}\right|^2 }{\left|{}^{Y_n^k}\bra{b_{n+1}} L_{d_{n}}^{Y_n^k} \ket{a_n}^{Y_n^k}\right|^2} \frac{p(b_{n+1}|Y_{n}^{k})}{p(a_{n}|Y_{n}^{k})}\right\} \\
&= \exp\left(-\beta \sum_{n=1}^{N} \Delta_{d_{n}}^{Y_n^k} \right)\frac{p(b_1)}{p(c_{N+1})}\  e^{-i_{\rm FB}^k [\psi_N]}\\
&= e^{-\sigma[\psi_N]-i_{\rm FB}^k[\psi_N]}.
\end{align*}
Thus, we can show that the inverse process introduced here satisfies both of the requirements (\ref{eq:GDFT}) and (\ref{eq:prob_norm}).

Finally, we discuss the sufficient condition for the absolute irreversibility term $\lambda_{\rm irr}$ to be zero. 
Since the absolute irreversibility is the consequence of the trajectories $\psi_N$ which satisfies $P_{k,\rm c}[\psi_N]=0$ and $\Tilde{P}_{k,\rm c} [\psi_N]\neq 0$ (i.e., $\psi_N \in (\mathcal{V}\smallsetminus\mathcal{W})$), as shown in Eq.~(\ref{eq:abs_irr}), the non-existence of such trajectories is a sufficient condition for $\lambda_{\rm irr} = 0$. Therefore, we can show that the following two equations are a sufficient condition for $\lambda_{\rm irr} =0$:
\begin{alignat}{2}
    p(b_1) \neq 0 &\ \ \ & &\mathrm{for\ all}\ b_1, \label{eq:abs_irr_suff_1}\\
    p(a_{n}|Y_{n}^{k})P[Y_{n}^{k}] \neq 0& \ \ \ & &\mathrm{for\ all}\ n, \ a_n, \ \mathrm{and}\ Y_n^k. \label{eq:abs_irr_suff_2}
\end{alignat}
From Eq.~(\ref{eq:abs_irr_suff_2}), we can show that $\Tilde{P}_{k,\rm c}[\psi_N]$ is always represented in the form of Eq.~(\ref{eq:prob_ref}). By comparing it with the definition of $P_{k,\rm c}[\psi_N]$, given by Eq.~(\ref{eq:prob_forward}), we can confirm there is no trajectory satisfying both $P_{k,\rm c}[\psi_N]=0$ and $\Tilde{P}_{k,\rm c} [\psi_N]\neq 0$, when Eqs.~(\ref{eq:abs_irr_suff_1}) and (\ref{eq:abs_irr_suff_2}) are satisfied.

\section{Comparison to the Framework Based on Quantum Causal Models}
In a series of previous works~\cite{strasberg2017quantum,strasberg2019operational,strasberg2019stochastic,strasberg2019repeated,strasberg2020thermodynamics,strasberg2022quantum}, a general framework of stochastic thermodynamics based on quantum causal models was considered, which is relevant to our setup of iterative quantum feedback.
In this section, we compare their framework based on quantum causal models with our theoretical framework used in this work, and discuss the differences in the conceptual approaches and the resulting thermodynamic laws. Specifically, we focus on the second law and the thermodynamic efficiency introduced in Ref.~\cite{strasberg2019operational}.

In Sec.~\ref{Ss:setup_compare}, we briefly review the thermodynamic framework of quantum causal models in the setup of quantum iterative feedback. In Sec.~\ref{Ss:SL_compare}, we compare the second law of thermodynamics derived in both frameworks. In Sec.~\ref{Ss:eff_compare}, we discuss the differences in the definitions of thermodynamic efficiency adopted in the two frameworks.

\subsection{Thermodynamic framework of quantum causal models} \label{Ss:setup_compare}

In the framework of quantum causal models, thermodynamics under interventions (such as measurement and feedback) is formulated by explicitly introducing ancilla systems (e.g., memories) that interact with the main system.
While this framework covers various stochastic dynamics, we here focus on the setup of iterative quantum feedback described in Appendix~\ref{appss:setup}.

For such setup, the state of the composite system, consisting of the main system and memories, after the $n$-th measurement is given by
\begin{equation}
    P[Y_n] \tau_{SM_n}^{Y_n} =  \mathcal{M}_{n}^{y_n} \circ \cdots  \circ \mathcal{L}_{2}^{Y_2} \circ \mathcal{M}_{2}^{y_2} \circ \mathcal{L}_{1}^{y_1} \circ \mathcal{M}_{1}^{y_1} (\rho_1 \otimes \ket{0}\bra{0}_{m_1} \otimes \cdots \otimes \ket{0}\bra{0}_{m_n}),
\end{equation}
and the state after the $n$-th feedback is defined as $\rho_{SM_n}^{Y_n} = \mathcal{L}_{n}^{Y_n} (\tau_{SM_n}^{Y_n})$.
Here, $\ket{0}_{m_n}$ denotes the initial state of the $n$-th memory. The map $\mathcal{L}_{n}^{Y_n}$ is the quantum channel corresponding to the $n$-th feedback (defined in Appendix~\ref{appss:setup}), which acts only on the system. The map $\mathcal{M}_{n}^{y_n}$ is the CP map corresponding to the $n$-th measurement with outcome $y_n$, defined by
\[
\mathcal{M}_{n}^{y_n}(\rho \otimes \ket{0}\bra{0}_{m_n}) \equiv \Pi_{y_n} V_n (\rho \otimes \ket{0}\bra{0}_{m_n}) V_n^\dag \Pi_{y_n},
\]
where $\Pi_{y_n}$ is a projector acting on the memory.
Specifically, the measurement with Kraus operators $\{M_{y_n,z_n}\}_{y_n,z_n}$ is realized by choosing $V_n$ and $\Pi_{y_n}$ such that
$M_{y_n,z_n} = {}_{m_n}\bra{y_n,z_n} V_n \ket{0}_{m_n}$ and $\Pi_{y_n} = \sum_{z_n} \ket{y_n,z_n} \bra{y_n,z_n}_{m_n}$, where $\{\ket{y_n,z_n}_{m_n}\}_{y_n,z_n}$ forms an orthonormal basis of the $n$-th memory.

In this setup, Ref.~\cite{strasberg2019operational} derives the following second-law-like inequalities for the $n$-th measurement and feedback processes, respectively:
\begin{align}
    \Sigma_{n,\rm M}^{Y_{n-1}} &\equiv S\left(\sum_{y_n} P[y_n|Y_{n-1}] \tau_{SM_n}^{Y_n}\right) - S\left(\rho_{SM_{n-1}}^{Y_{n-1}}\right) \geq 0,  \label{seq:meas_SL_comp}\\
    \Sigma_{n,\rm FB}^{Y_{n}} &\equiv S(\rho_{SM_n}^{Y_n}) -S(\tau_{SM_n}^{Y_n}) - \beta Q_n^{Y_n} \geq 0,\label{seq:FB_SL_comp}
\end{align}
where $\Sigma_{n,\rm M}^{Y_{n-1}}$ and $\Sigma_{n,\rm FB}^{Y_{n}}$ denote the conditional entropy production of the total system during the $n$-th measurement and feedback processes, respectively. Here, $Q_n^{Y_n}$ is the heat absorbed from the bath in the $n$-th feedback process, conditioned on the measurement outcomes $Y_n$.
These expressions are defined with respect to the composite system of the main system and the memories, which differs conceptually from the second laws derived in our framework. In general, when there exists an unmonitored label $z_n$, the conditional states $\rho_{SM_n}^{Y_n}$ and $\tau_{SM_n}^{Y_n}$ include nontrivial quantum correlations between the system and the memories. Therefore, the quantities such as $S(\rho_{SM_n}^{Y_n})$ and $S(\tau_{SM_n}^{Y_n})$ cannot be represented solely with the system state. This makes a direct comparison with the second law derived in this work challenging.

However, in the special case where no unmonitored labels $z_n$ are present, the projectors $\Pi_{y_n}$ become rank-one, and the composite conditional states reduce to product forms:
\begin{align}
    \tau_{SM_n}^{Y_n} &= \tau^{Y_n}_n \otimes \Pi_{y_n} \otimes \cdots \otimes \Pi_{y_2} \otimes \Pi_{y_1}, \label{eq:tau_comp}\\
    \rho_{SM_n}^{Y_n} &= \rho^{Y_n}_{n+1} \otimes \Pi_{y_n} \otimes \cdots \otimes \Pi_{y_2} \otimes \Pi_{y_1}. \label{eq:rho_comp}
\end{align}
In this case, the inequalities~\eqref{seq:meas_SL_comp} and \eqref{seq:FB_SL_comp} can be simplified and expressed solely in terms of the system’s conditional states.
In Secs.~\ref{Ss:SL_compare} and~\ref{Ss:eff_compare}, we compare these simplified forms of the second laws and their associated thermodynamic efficiencies.

\subsection{Second law of thermodynamics} \label{Ss:SL_compare}

In the situation where the composite system is described as in Eqs.~\eqref{eq:tau_comp} and~\eqref{eq:rho_comp}, the second laws for the composite system given in Eqs.~\eqref{seq:meas_SL_comp} and~\eqref{seq:FB_SL_comp} can be expressed solely in terms of the system state as follows:
\begin{align}
    \Sigma_{n,\rm M}^{Y_{n-1}} &= H(y_n|Y_{n-1}) + \sum_{y_n} P[y_n|Y_{n-1}] S\left( \tau^{Y_n}_n \right) - S\left( \rho^{Y_{n-1}}_n \right) \geq 0, \\
    \Sigma_{n,\rm FB}^{Y_{n}} &= S(\rho^{Y_n}_{n+1}) - S(\tau^{Y_n}_n) - \beta Q_n^{Y_n} \geq 0, \label{seq:SL_noz_compare}
\end{align}
where $H(a|b) \equiv \sum_{a,b} p(a,b) \ln \frac{p(b)}{p(a,b)}$ is the conditional Shannon entropy.

In particular, the second law for the feedback process, Eq.~\eqref{seq:SL_noz_compare}, can be directly related to the second law derived in our framework. Taking the ensemble average of Eq.~\eqref{seq:SL_noz_compare} yields
\begin{equation}
    \sum_{n} \sum_{Y_n} P[Y_n] \Sigma_{n,\rm FB}^{Y_{n}} = \langle \sigma \rangle + \langle i_{\rm QCT} \rangle  - \chi (\rho_{N+1} : Y_N) \geq 0, \label{seq:SL_nonz_stras}
\end{equation}
where the entropy production $\sigma$, QC-transfer entropy $i_{\rm QCT}$, and the Holevo information $\chi$ are defined in Appendix~\ref{apps:thermo_QCTE}. This reproduces the generalized second law derived in Ref.~\cite{yada2022quantum}. (Note that this correspondence is not valid when unmonitored label $z_n$ exists.)

On the other hand, the second law derived in our work, Eq.~\eqref{appeq:GSL_BQC}, namely $\langle \sigma \rangle + \langle i_{\rm FB}^k \rangle \geq 0,$ provides a tighter bound than Eq.~\eqref{seq:SL_nonz_stras}, because the inequality $\langle i_{\rm FB}^k \rangle \leq \langle i_{\rm QCT} \rangle  - \chi (\rho_{N+1} : Y_N)$ always holds (see the derivation of Eq.~\eqref{eq:BQCTE_size}):
\begin{equation}
    \sum_{n} \sum_{Y_n} P[Y_n] \Sigma_{n,\rm FB}^{Y_{n}} \geq \langle \sigma \rangle + \langle i_{\rm FB}^k \rangle \geq 0.
\end{equation}
This is because the second law in our framework provides a more precise bound that reflects the causal structure of the feedback, whereas the framework in Ref.~\cite{strasberg2019operational} is designed to be broadly applicable without requiring any details of the feedback protocols.

Furthermore, Ref.~\cite{strasberg2019operational} also derives a second law for the entire $N$-cycle process by taking the ensemble average over all $Y_N$, as follows:
\begin{equation}
    \Sigma_{\rm tot} \equiv \sum_{n=1}^N \left[\sum_{Y_n} P[Y_n] \Sigma_{n,\rm FB}^{Y_{n}} + \sum_{Y_{n-1}}  P[Y_{n-1}]  \Sigma_{n,\rm M}^{Y_{n-1}}\right] = H(Y_N) + \sum_{Y_N} P[Y_N] S(\rho_{N+1}^{Y_N}) - S(\rho_1) - \beta \langle Q\rangle \geq 0, \label{seq:GSL_stras}
\end{equation}
where $\Sigma_{\rm tot}$ represents the total entropy production of the composite system, and $H(y) \equiv \sum_y -p(y) \ln p(y)$ denotes the Shannon entropy. This form of the second law will be essential in defining the thermodynamic efficiency $\zeta$ discussed in the next subsection.

\subsection{Thermodynamic efficiency} \label{Ss:eff_compare}

We next discuss the relationship between the efficiency $\eta$ introduced in Eq.~\eqref{eq:eff} in our work and the efficiency discussed in Ref.~\cite{strasberg2019operational}. It is defined based on the second law~\eqref{seq:GSL_stras}, where the total entropy production can be decomposed as
\begin{equation}
    \Sigma_{\rm tot} = H(Y_N) + \beta (\langle W \rangle - \langle \Delta f \rangle).
\end{equation}
Here, the average values of the work input $\langle W \rangle$ and the conditional free energy change $\langle \Delta f \rangle$ are given by:
\begin{align}
    \langle W \rangle &= \tr[H_{\rm S} (\rho_{N+1} - \rho_1)]  - \langle Q\rangle,\\
    \langle \Delta f \rangle &= \tr[H_{\rm S} (\rho_{N+1} - \rho_1)] - k_B T \left[\sum_{Y_N} P[Y_N] S(\rho_{N+1}^{Y_N}) - S(\rho_{1}) \right],
\end{align}
where $H_{\rm S}$ represents the system Hamiltonian, and the first term in both $\langle W \rangle$ and $\langle \Delta f \rangle$ corresponds to the energy change of the system during the $N$-cycle process.
We note that $\Delta f $ is the free energy change conditioned on all past measurement outcomes. Based on this decomposition and the inequality $\Sigma_{\rm tot} \geq 0$, the efficiency is defined as~\cite{strasberg2019operational}:
\begin{equation}
    \zeta \equiv \frac{\langle \Delta f \rangle}{\langle W \rangle + k_B T H(Y_N)},
\end{equation}
which quantifies the rate that the work input and the entropy of the measurement outcomes are converted into the conditional free energy change.

On the other hand, the efficiency introduced in this work is defined as
\begin{equation}
\label{eq:eff_our}
    \eta \equiv \frac{-\langle \sigma \rangle}{\langle i_{\rm QCT} \rangle} = \frac{-\beta(\langle W \rangle - \Delta F)}{\langle i_{\rm QCT} \rangle},
\end{equation}
where $\Delta F = \tr[H_{\rm S} (\rho_{N+1} - \rho_1)] - k_B T \left[S(\rho_{N+1}) - S(\rho_1) \right]$ is the free energy change (not conditioned on measurement outcomes).
Both the denominator and numerator of the efficiency $\eta$ have clear operational interpretations. The denominator $\langle i_{\rm QCT} \rangle$ quantifies the total information gain through iterative quantum measurements, as discussed in the main text. The numerator $-\langle \sigma \rangle = -\beta(\langle W \rangle - \Delta F)$ corresponds to the thermodynamic benefit, which is non-positive in the absence of feedback control. In many situations, this quantity serves as the target to be maximized by feedback control.
For example, in our experiment, where the system’s inverse temperature is $\beta = 0$, it is reduced to the entropy reduction $-\langle \sigma \rangle = -\Delta S$, which directly reflects the stability of the target state. Furthermore, in scenarios where $\Delta F = 0$ (e.g., steady states or cyclic quantum heat engines), the numerator corresponds to the extractable work.
Thus, the efficiency $\eta$ characterizes how effectively the acquired information is converted into thermodynamic benefit, and is conceptually aligned with measures of information-thermodynamic efficiency widely used in the previous experimental works~\cite{toyabe2010experimental,koski2014experimental,masuyama2018information}.

\section{Equivalent description of the dynamics}

\begin{figure*}[]
\begin{center}
\includegraphics[width=0.7\textwidth]{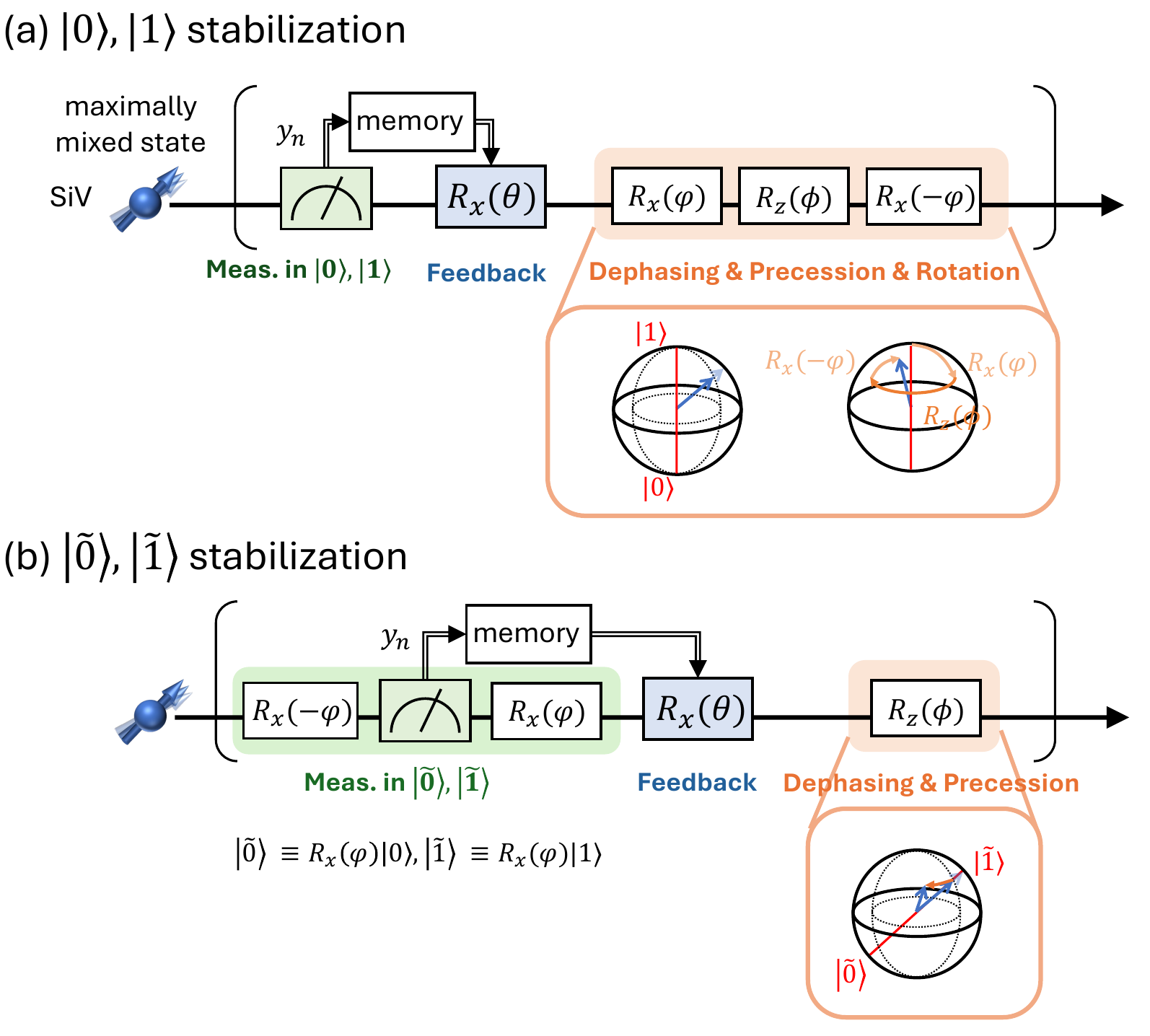}
\caption{Equivalent dynamics with different interpretations.  
(a) Schematic of the iterative feedback protocol where the measurement is performed in the $\sigma_z$ basis and dephasing occurs in a tilted basis. This is the description adopted in the current manuscript.  
(b) Schematic of the feedback protocol where the measurement is performed in a tilted basis $\{\ket{\widetilde{0}}, \ket{\widetilde{1}}\}$ and dephasing occurs in the $\sigma_z$ basis.  
These two dynamics are mathematically equivalent.
}
\label{fig:equiv_def}
\end{center}
\end{figure*}

In this section, we explain that our experiment, which has been described in the previous sections as (a) measurement in the $\{\ket{0}, \ket{1}\}$ basis followed by the decoherence in the tilted basis (with external rotation $R_x(\pm\varphi)$), can be equivalently interpreted as (b) measurement in the tilted basis $\{\ket{\widetilde{0}}, \ket{\widetilde{1}}\}$  followed by decoherence with $\{\ket{0}, \ket{1}\}$ basis.
This equivalence is illustrated in Fig.~\ref{fig:equiv_def}.
In the new interpretation (b), the rotations $R_x(\pm\varphi)$ are understood as modifying the measurement basis, where $\ket{\widetilde{0}}\equiv R_x(\varphi)\ket{0}$ and $\ket{\widetilde{1}}\equiv R_x(\varphi)\ket{1}$. The Kraus operators for the new measurement process become $\{R_x(\varphi) M_{y_n,z_n} R_x(-\varphi)\}_{y_n,z_n}$.  Since the measurable basis in the SiV center experiment is fixed to the $\sigma_z$ basis, implementing the rotations $R_x(\pm\varphi)$ before and after the measurement is the standard way to effectively realize a measurement in a tilted basis.
In this interpretation, the artificial rotations $R_x(\pm\varphi)$ are absorbed into the measurement process, and the entire protocol can be viewed as feedback control in the tilted basis against spontaneous decoherence in the $\{\ket{0}, \ket{1}\}$ basis.

\end{document}